\documentclass[fleqn,usenatbib]{mnras}
\usepackage{newtxtext,newtxmath}
\usepackage[T1]{fontenc}
\usepackage{ae,aecompl}
\usepackage{graphicx}	
\usepackage{amsmath}	
\usepackage{amssymb}	
\usepackage{multirow}
\usepackage{xcolor}
\usepackage{colortbl}
\usepackage{gensymb}
\usepackage{soul}
\usepackage{diagbox}

\def\zs{\mbox{{$z_{\rm spec}$}}}
\def\zp{\mbox{{$z_{\rm phot}$}}}

\newcommand{\zspec}{z$_{\mbox{spec}}$}

\newcommand*\mathinhead[2]{\texorpdfstring{$\boldsymbol{#1}$}{#2}}
\definecolor{cadmiumred}{rgb}{0.89, 0.0, 0.13}
\newcommand{\red}[1]{\textcolor{cadmiumred}{\textbf{#1}}}
\definecolor{mygreen}{RGB}{0, 200, 0}
\definecolor{mygrey}{gray}{0.6}
\definecolor{scc}{rgb}{1.0, 0.49, 0.0}
\defcitealias{Ananna2017}{A17}
\title[Photo-z for X-ray selected  AGN in the eROSITA era]{Photometric Redshifts for X-ray selected Active Galactic Nuclei in the eROSITA era}

\author[M. Brescia et al.]{
M. Brescia$^{1}$\thanks{E-mail: massimo.brescia@inaf.it},
M. Salvato$^{2\thanks{E-mail: mara@mpe.mpg.de}}$,
S. Cavuoti$^{1,3,4\thanks{E-mail: stefano.cavuoti@gmail.com}}$,
T.~T. Ananna$^{5,6}$,
G. Riccio$^{1}$, \and 
S.~M. LaMassa$^{7}$,
C.~M. Urry$^{6}$ and
G. Longo$^{3,4}$
\\
$^{1}$INAF - Astronomical Observatory of Capodimonte, via Moiariello 16, I-80131, Napoli, Italy\\
$^{2}$Max-Planck-Institut f{\"u}r extraterrestrische Physik, Postfach 1312, Giessenbachstra{\ss}e, D-85741 Garching, Germany\\
$^{3}$Department of Physics ``E. Pancini'', University Federico II, via Cinthia 6, I-80126, Napoli, Italy\\
$^{4}$INFN section of Naples, via Cinthia 6, I-80126, Napoli, Italy\\
$^{5}$Department of Physics, Yale University, P.O. Box 201820, New Haven, CT 06520-8120, USA\\
$^{6}$Yale Center for Astronomy and Astrophysics, P.O. Box 208121, New Haven, CT 06520, USA\\
$^{7}$Space Telescope Science Institute, 3700 San Martin Drive, Baltimore, MD 21218, USA
}
\date{Accepted XXX. Received YYY; in original form ZZZ}

\pubyear{2019}

\begin{document}
\label{firstpage}
\pagerange{\pageref{firstpage}--\pageref{lastpage}}
\maketitle

\begin{abstract}
With the launch of eROSITA, successfully occurred  on July 13th, 2019, we are facing the challenge of computing 
reliable photometric redshifts for 3 million of AGNs over the entire sky, having available only patchy and inhomogeneous 
ancillary data. While we have a good understanding of the photo-z quality obtainable for AGN using SED-fitting technique, 
we tested the capability of Machine Learning (ML), usually reliable in computing photo-z for QSO in wide and shallow areas 
with rich spectroscopic samples. Using MLPQNA as example of ML, we computed photo-z for the X-ray selected sources in Stripe 
82X, using the publicly available  photometric and spectroscopic catalogues. Stripe 82X is at least as deep as eROSITA will 
be and wide enough to include also rare and bright AGNs. In addition, the availability of ancillary data mimics what can be 
available in the whole sky. We found that when optical, NIR and MIR data are available, ML and SED-fitting perform comparably 
well in terms of overall accuracy, realistic redshift probability density functions and  fraction of outliers, although they 
are not the same for the two methods. The results could further improve if the photometry available is accurate and including 
morphological information. Assuming that we can gather sufficient spectroscopy to build a representative training sample, with 
the current photometry coverage we can obtain reliable photo-z for a large fraction of sources in the Southern Hemisphere well 
before the spectroscopic follow-up, thus timely enabling the eROSITA science return. The photo-z catalogue is released here.
\end{abstract}


%%%%%%%%%%%%%%%%%%%%%%%%%%%%%%%%%%%%%%%%%%%%%%%%%%

%%%%%%%%%%%%%%%%% BODY OF PAPER %%%%%%%%%%%%%%%%%%

\section{\label{sec:Intro} Introduction}
Photometric redshifts (photo-z) are now routinely used in many applications, from galaxy evolution to cosmological studies. In particular, 
the present and planned photometric surveys over  many thousand square degrees (e.g., DES, Euclid, LSST, eROSITA, SpherEx) rely  mostly on 
photo-z for their scientific exploitation.
As known, there are basically two classes of methods commonly used to derive photo-z: the template Spectral Energy Distribution (SED) fitting 
methods \citep[e.g.,][]{Bolzonella00, Ilbert06, Tanaka15} and the empirical or interpolative methods \citep[e.g.,][]{Cavuoti15a,Carrasco13}. 
Both  methods are characterized by advantages and shortcomings \citep[for a complete review see][]{Salvato18}, but essentially both rely on colour/magnitude-redshift maps, with the difference that SED based methods assume {\it a priori} knowledge of the map, while empirical methods, based on Machine Learning (ML), learn the map anew from the data every time. Recently more algorithms that merge the pros of the two techniques are developed and show very promising results also in computing photo-z for mixed populations of galaxies and AGNs \citep[e.g.,][]{Duncan18}, in some case being also able to successfully characterize the sources \citep{Fotopoulou18}.

SED fitting techniques are able to provide all at once photo-z point estimates, photo-z Probability Density Function (PDZ) and the spectral type of each source, at any redshift.

In the supervised ML techniques, the learning process is regulated by the spectroscopic information (i.e. redshift) available for a sub-sample of the objects. ML methods -- ideal for  the million of sources provided by multi-wavelength surveys -- are extremely precise as long as the spectroscopic sample is representative of the population for which the photo-z has to be computed. 
This means that they cannot provide accurate solutions outside z-spec range of the training set. 

Even though the accuracy reached by empirical methods and SED-fitting for inactive galaxies are comparable, is not the case for galaxies hosting AGNs. For these sources in fact, the amount of contribution from the AGN to the total emission in the various bands is {\it a priori} unknown. 
In photo-z computed via SED-fitting this translates in the difficulty of defining for every survey the correct set of templates forming the library \citep[e.g.,][hereafter \citetalias{Ananna2017}]{Salvato09,Cardamone10, Luo10, Salvato11, Hsu14, Ananna2017}. Similarly, in photo-z computed with empirical methods, it is necessary to have a very large and complete spectroscopic sample to use as training set. That is why in the past, there have been only few attempts to compute photo-z for AGN with ML based methods \citep[e.g.][]{Budavari01, Bovy12, Brescia13}. 
All these pioneering works focused on the SDSS footprints (where the plethora of spectroscopic redshifts, hereafter \zspec, are well 
suited to ML techniques) and on optically selected QSOs, mostly at z $\ge 1.5$, which are dominated by the AGN, with little contribution from the host.
For low redshift and low luminosity AGN (i.e. Seyfert galaxies), where the host galaxy contribution to the total emission is significant, both the accuracy and fraction of outliers of the photo-z computed with SED-fitting technique are comparable to the results for normal galaxies ($\sigma_{\rm NMAD}\sim$ 1\%; $\eta \sim$5\%). However this is true only in the fields where narrow and/or intermediate filter band photometry is available, like e.g, in COSMOS, CDFS, Alhambra \citep[e.g.][]{Salvato09, Salvato11, Marchesi16, Cardamone10, Matute12}. In fact, the narrow/intermediate band photometry easily pinpoints the emission lines, typical SED features for AGN.
For a mixed set of AGNs and with only broad band photometry, even with coverage from UV to MIR, photo-z for AGN via SED-fitting reach an accuracy of about 6-8\% with about 
18-25\% in fraction of outliers \citep[e.g.,][ \citetalias{Ananna2017}]{Fotopoulou12, Nandra15}, with the accuracy of the extended sources better than those classified as point-like in optical images. The situation gets even more difficult when the survey is wide and the photometry is assembled from heterogeneous photometric catalogues rather than computed in a consistent way from homogenized images \citep[e.g., COSMOS]{Laigle16, Ilbert09}. 

ML methods are less affected by this problem because they account for the differences in the photometry defined in the various catalogues, but the photo-z computation must then be preceded by a search for the best features e.g., certain type of magnitudes, or specific photometric bands/colors \citep[e.g., see][]{Polsterer14, D'Isanto18, Fotopoulou18, Ruiz18}. In addition, the quality degrades fast, when not sufficient photometry is available.

Photo-z can replace spectroscopic redshifts in AGN evolution studies or AGN clustering only if accompanied by the redshift probability density function \citep[PDZ; e.g.,][]{Miyaji15, Georgakakis14}. While PDZs are routinely produced when computing photo-z via SED-fitting, their production with empirical methods requires an additional computational effort that only recently became more feasible \cite[e.g.,][]{Sadeh16, Cavuoti17, Amaro18, Brescia18, Mountrichas17, Duncan17}.

In terms of computational speed, ML outperforms the template-fitting techniques \citep[][]{Vanzella04}. 
This makes ML a natural choice for the computation of photo-z for the very large forthcoming deep and wide surveys such as Euclid 
\citep[][]{Laureijs2010} and LSST \citep[][]{Ivezic19}, in which the computation of photo-z will be a real challenge. 
For AGN, the next challenge is presented by the $\sim3$ million sources that eROSITA \citep[extended Roentgen Survey with an Imaging Telescope 
Array;][]{Merloni12}, the primary instrument on the Russian Spektrum-Roentgen-Gamma (SRG) mission, will detect. eROSITA will provide an all-sky X-ray survey every $6$ months for $4$ years, with a  final expected depth of $1\times10^{-14} {\rm erg/cm^2/s}$ ($3\times10^{-15} {\rm erg/cm^2/s}$ at the poles) which is about 30 times deeper than ROSAT \citep[][]{Voges99, Boller16} in the soft band (0.5-2 keV). eROSITA will also provide for the  first time ever an all-sky image in the hard band (2-10 keV), reaching an expected depth of $2\times10^{-13} {\rm erg/cm2/s}$ ($4\times10^{-14} {\rm erg/cm2/s}$ at the poles). 
With this depth, eROSITA will detect the low luminosity AGN that are present in deep pencil-beam surveys, but also the bright and more rare objects that are observed in wide areas. 

The recent release by \citetalias{Ananna2017} of the complete photometry of the counterparts to the X-ray sources detected in Stripe 82X \citep[][]{LaMassa16, LaMassa13b, LaMassa13} offers the possibility to study the performances of photo-z for AGN via ML in a complete way. Stripe 82X covers an area of about $31$ deg$^2$, with a X-ray depth of $8.7\times10^{-16}$ erg/cm$^2$/s in the soft band and $4.7\times10^{-15}$ erg/cm$^2$/s in the hard band and it includes about $6,000$ sources, $\sim$ $3000$ of which are provided with reliable zspec from SDSS.
The photometric catalogue includes Galex, SDSS, UKIRT, VHS, SPITZER/IRAC and WISE with a depth sufficient to detect the X-ray sources at least at the depth of eROSITA. Thus, Stripe 82X can mimic eROSITA in terms of X-ray depth and ancillary data coverage that will be available in the whole sky, thanks to the coverage provided by e.g., PanStarrs, skyMapper, DES, VHS, UKIDSS, WISE (Spitzer/IRAC will be available only for patches of $100$ deg$^2$) and with LSST and SpherEx in the future.
\citetalias{Ananna2017} provides not only photometry, but also photo-z and related PDZs computed via SED fitting using LePhare \citep{Arnouts99, Ilbert06}, after splitting the sample in subgroups, each fitted with a dedicated library of templates.

The work presented in this paper consists of two parts.
In the first part we evaluate and optimise the space of the parameters that improve the accuracy of ML. This is done by selecting the best features \citep{Lal06}. 
In the literature \citep[e.g.,][]{Guyon03}, there are plenty of algorithms aimed at the selection of the best features, such as Principal Component Analysis \citep[PCA;][]{Jolliffe02}, \textit{filter} techniques, based on the evaluation of single features through a variety of significance tests, generally fast but less accurate \citep{Gheyas10}, \textit{wrapper} methods, which make use of an arbitrary learning algorithm (such as neural networks or nearest-neighbour) to evaluate the relevance of feature sets \citep{Kohavi97} and \textit{embedded} methods, performing a feature selection during the prediction/classification model training procedure (a typical example is the Random Forest model, \citealt[]{Breiman01}).
Here we use a novel feature selection method, named $\Phi$LAB (PhiLAB, Parameter handling investigation LABoratory), a hybrid approach that includes properties of both wrappers and embedded feature selection categories \citep{delliveneri19}.
For this purpose, the determination of the features, rather than the computation of the photo-z, is relevant. Because the best features are not available for the entire sample, our purpose here is to list them, for their use in other surveys.
The computation of the photo-z via ML is discussed in the second part of the paper and for the reason described above it will be computed for the entire sample also in a more traditional way using MLPQNA \citep[][]{Brescia13, Cavuoti12, Cavuoti15}. Missing some of the best features, will translate in the degradation of the accuracy.

For each of the sources we provide also the PDZ through METAPHOR \citep{Cavuoti17} and we will compare the photo-z and the PDZ with the results presented in \citetalias{Ananna2017}. In particular we will test the accuracy of the new photo-z at the depth of eROSITA all-sky survey.

\noindent \textbf{Outline:} In Sec.~\ref{sec:Data} we describe the data used in this work while in Sec.~\ref{sec:Method} we briefly describe the methods involved in our analysis.
In Sec.~\ref{sec:ps} we analyze the parameter space and its optimization, while in Sec.~\ref{sec:tomo} we discuss the impact of X-Ray flux, photometry and morphology in the quality of photo-z.
The Sections ~\ref{sec:pointEst} and ~\ref{sec:pdzEst} are devoted to the analysis of photo-z and PDZ estimation, respectively.
Finally the conclusions are drawn in Sec.~\ref{sec:Conclusions}.
In the Appendix~\ref{app:catalog}, the catalogue of photo-z computed with MLPQNA and used for this work are made publicly available, 
while PDZs are available under request.\\
In this paper, unless differently stated, we use magnitude expressed in AB and adopt a cosmology of H$_0$ = $70$ km s$^{-1}{\rm Mpc}^{-1}$, $\Omega_M$ = $0.27$, and $\Lambda$ = $0.73$.\\

\section{Data}\label{sec:Data} 
This section is dedicated to describe all photometric and spectroscopic data used in the experiments.

\subsection{PHOTOMETRY}\label{sec:photo}
\begin{figure*}
\centering
\includegraphics[]{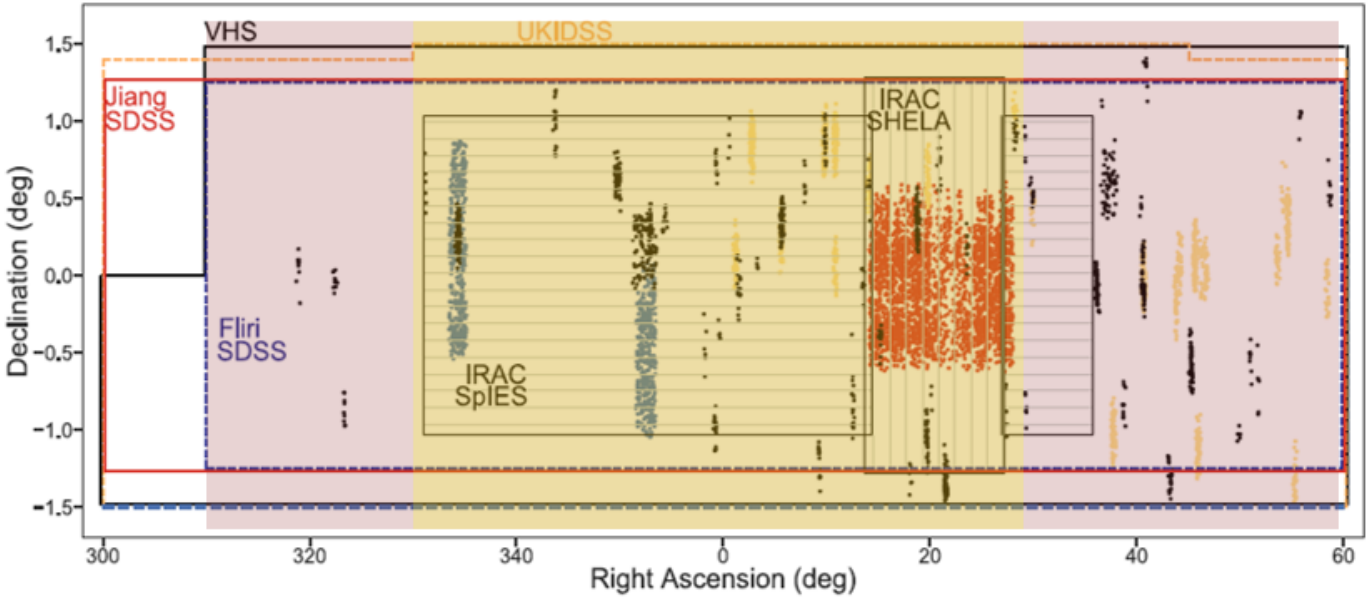}
\caption{Map of the original multi-wavelength coverage of Stripe 82X area discussed in \citetalias{Ananna2017}. The total area extends for $\sim 2.5\degree$ in Declination and $120\degree$ in Right Ascension. The dots represent X-ray sources, respectively, from XMM-Newton AO13 (red), AO10 (blue), archival XMM-Newton sources (yellow) and Chandra sources (black). While standard photo-z are generated for the entire area (in red), the selection of the best features discussed in the first part of the paper is obtained considering only the sources in the yellow area.}
\label{fig:radec}
\end{figure*}

The photometry used in this work is extracted from the catalogue presented in \citetalias{Ananna2017}, which lists the multi-wavelength properties of the counterparts to the  X-ray sources detected in Stripe 82X. Compared to the previous version of the catalogue presented in \cite{LaMassa16}, this new catalogue uses deeper multi-wavelength data for the identification of the counterparts and for the computation of the photometric redshifts via SED fitting. Although the catalogue of \citetalias{Ananna2017} includes $6,187$ X-ray sources, we focus here only on the $5,990$ for which a reliable counterpart was identified.
All the details on the properties of the photometric dataset are exhaustively presented in \citetalias{Ananna2017}. Here we provide only the list of data that we use for this paper, focusing on the central area of Stripe 82X, observed by SPITZER, as shown in Fig.~\ref{fig:radec}.
In particular we considered:
\begin{itemize}
\item FUV and NUV magnitudes and corresponding errors from GALEX all-sky survey \citep[]{Martin05}. They were not used in this work due 
to the shallowness of the data;
\item  {\it u,g,r,i,z}  SDSS AUTO magnitudes and corresponding errors from \citet[][]{Fliri16};
\item  J, H, K 
from VISTA \citep[]{Irwin04}. As shown in \citetalias{Ananna2017} additional data in J$_{\rm UK}$,H$_{\rm UK}$,K$_{\rm UK}$ data from UKIDSS \citep[]{Lawrence07} are available for the same area  but were not used in this paper;

\item  $3.6$ and  $4.5~\mu$m magnitudes and corresponding errors from IRAC. Here two complementary surveys are used: SPIES \citep[][]{Timlin} and SHELA \citep{Papovich}. Given the similarity of the two surveys, we do not differentiate sources belonging to one or another;
\item W1, W2, W3, W4 magnitudes and corresponding errors from AllWISE \citep{Wright10}.
\end{itemize}
In the first column of Table \ref{tab:depth}, we report the nominal depth of each photometric band considered. 

The original catalogue is complemented by Soft, Hard and Full band X-ray fluxes from {\it XMM-Newton} and {\it Chandra} \citep[see][for details]{LaMassa16}. It also includes morphological information on the extension and variability of the sources in the optical band. We retain such information, as it has been already demonstrated in literature that they affect the accuracy of photo-z for AGN and can be used as priors for improving performance \citep[e.g.,][]{Salvato09}. While these data are not used directly for the computation of the photo-z, they are employed to perform various experiments by creating sub-samples in X-ray flux and morphology.

%%%%%%%%%%%%%%%%%%%%%%%%%%%%%%%%%%%%%%%%%%%%%%%%%
%%%%%%%%%%%%%%%%%%%%%%%%%%%%%%%%%%%%%%%%%%%%%%%%%%%%%%%%%%%%%%

\begin{table*}
\centering  \resizebox{\textwidth}{!}{  

\begin{tabular}{lcccccccccr}
Filter &	\multicolumn{9}{c}{BAND DEPTH}\\
\hline
 & \multirow{2}{*}{NOMINAL} &\multirow{2}{*}{BEST} &\multirow{2}{*}{SDSS} & SDSS \& & SDSS \& & SDSS \& & SDSS        & SDSS        & SDSS VHS   \\
 &  		                & 	                   &		              & VHS	    & IRAC    & WISE	& VHS \& IRAC & VHS \& WISE & IRAC \& WISE\\

 \hline
 \hline
%	&	\multicolumn{2}{c}{		\\ \hline
%FUV			    &	21.99	& ---	                        & ---   & ---   & ---                           & ---   & ---                            &---    & ---\\
%NUV			    &	21.99	& ---	                        & ---   & ---   & ---                           & ---   & ---                            & ---   & --- \\
u			    &	31.22	& 28.54	                        & 28.54 & 28.54 & 28.54                         & 28.54 & 28.54                          & 28.54 & 28.54 \\
g			    &	28.77	& 24.20	                        & 24.39 & 24.20 & 24.39                         & 24.39 & 24.20                          & 24.20 & 24.20\\
r			    &	27.13	& 23.25	                        & 23.43 & 23.25 & 23.43                         & 23.43 & 23.25                          & 23.25 & 23.25\\
i			    &	27.21	& 22.35	                        & 23.49 & 22.64 & 23.49                         & 22.45 & 22.64                          & 22.35 & 22.35\\
z			    &	30.46	& 22.42	                        & 23.35 & 22.46 & 22.99                         & 22.42 & 22.46                          & 22.42 & 22.08\\
J			    &	24.74	& 21.64	                        & ---   & 24.64 & ---                           & ---   & 21.64                          & 21.64 & 21.51 \\
H			    &	24.15	& 22.87                         & ---   & 22.87 & ---                           & ---   & 21.61                          & 22.87 & 21.61\\
K			    &	22.60	& 21.63                         & ---   & 21.63 & ---                           & ---   & 21.63                          & 21.63 & 21.63\\
%Juk			    &	23.44	& ---                           & ---   & ---   & ---                           & ---   & ---                            & ---   & ---\\
%Huk			    &	22.69	& ---	                        & ---   & ---   & ---                           & ---   & ---                            & ---   & ---\\
%Kuk			    &	22.41	& ---	                        & ---   & ---   & ---                           & ---   & ---                            & ---   & ---\\
CH1\_SPIES	    &	24.27	& \multirow{2}{*}{20.82$^\dag$} & ---   & ---   & \multirow{2}{*}{21.64$^\dag$} & ---   & \multirow{2}{*}{21.06$^\dag$}  & ---   & \multirow{2}{*}{20.49$^\dag$} \\
CH1\_SHELA	    &	22.80	&                               & ---   & ---   &                               & ---   &                                & ---   & \\
CH2\_SPIES	    &	22.88	& \multirow{2}{*}{20.49$\dag$}  & ---   & ---   & \multirow{2}{*}{21.41$\dag$}  & ---   & \multirow{2}{*}{21.07$^\dag$}  & ---   & \multirow{2}{*}{20.22$^\dag$}\\
CH2\_SHELA	    &	23.88	&                               & ---   & ---   &                               & ---   &                                & ---   & \\
W1			    &	21.16	& 20.71                         & ---   & ---   & ---                           & 20.71 & ---                            & 20.71 & 20.61\\
W2			    &	20.74	& 20.59                         & ---   & ---   & ---                           & 20.63 & ---                            & 20.63 & 20.59\\
W3			    &	18.20	& 18.04                         & ---   & ---   & ---                           & 18.11 & ---                            & 18.11 & 18.04\\
W4			    &	16.15	& 16.06                         & ---   & ---   & ---                           & 16.13 & ---                            & 16.13 & 15.94\\
\hline
N. of sources	& 5990	    & 2290                          & 4855  & 3218  & 2293 & 3291  & 1620 & 2696  & 1380 \\
\hline
N. of sources   &	\multirow{2}{*}{2933}& \multirow{2}{*}{1686}&	\multirow{2}{*}{2793}	& \multirow{2}{*}{2218}	& \multirow{2}{*}{1596} &	\multirow{2}{*}{2160}	& \multirow{2}{*}{1279} & 	\multirow{2}{*}{1935}	&\multirow{2}{*}{1121}	\\ 
w/ z$_{\rm spec}$ &	& & & & & & & & \\
\hline
N. of sources &	\multirow{2}{*}{ 2351}& \multirow{2}{*}{1249 }&	\multirow{2}{*}{ 2025}	& \multirow{2}{*}{ 1649}	& \multirow{2}{*}{1051} &	\multirow{2}{*}{1619}	& \multirow{2}{*}{888} &	\multirow{2}{*}{1445}	&\multirow{2}{*}{793 }	\\ 
w/ F$_X>10^{-14}$ &	& & & & & & & & \\
\hline
N. of sources  & & & & & & & & & \\
w/ F$_X>10^{-14}$ & 1550& 1025& 1483& 1309 & 857 & 1256 & 758 & 1174 & 683 \\
and z$_{\rm spec}$& & & & & & & & & \\
%Num. of Bands	&	21	&	14\\
%max. redshift	&	5.3912	&	5.3912\\\hline
\hline\hline
\end{tabular}
}
\caption{Summary table for depth, amount of sources and redshift coverage.
The first column refers to the nominal depth of the entire sample of reliable counterparts in Stripe 82X, as presented in \citetalias{Ananna2017}. The following columns refer to the magnitudes reached in the various experiments, i.e., the faintest magnitude reported in the Stripe 82X catalogue for the various sub-samples for which the photo-z have been computed. The values in the column $BEST$ represent the faintest magnitudes of the sub-sample of sources in the yellow area of Fig.~\ref{fig:radec}, used for the features analysis performed with $\Phi$LAB, (Sec.~\ref{sec:philab}). The bands marked with a --- symbol have been discarded from that specific experiment.\newline$^\dag$ SPIES and SHELA have been used together (Sec.~\ref{sec:photo}). The last rows of the table list, for each of the sub-samples, the total number of sources, their amount with spectroscopic redshift available, those with a X-ray flux brighter than $10^{-14}$ $erg/cm^2/s$ and the final depth expected by eROSITA all-sky survey. Note that here we use only the sources for which the determination of the counterpart is secure, i.e. (SDSS,VHS,IRAC)\_REL\_CLASS==SECURE in the catalogue of \citetalias{Ananna2017}}
\label{tab:depth}
\end{table*}

\subsection{SPECTROSCOPY}
\begin{figure}
\centering
	\includegraphics[width=\columnwidth]{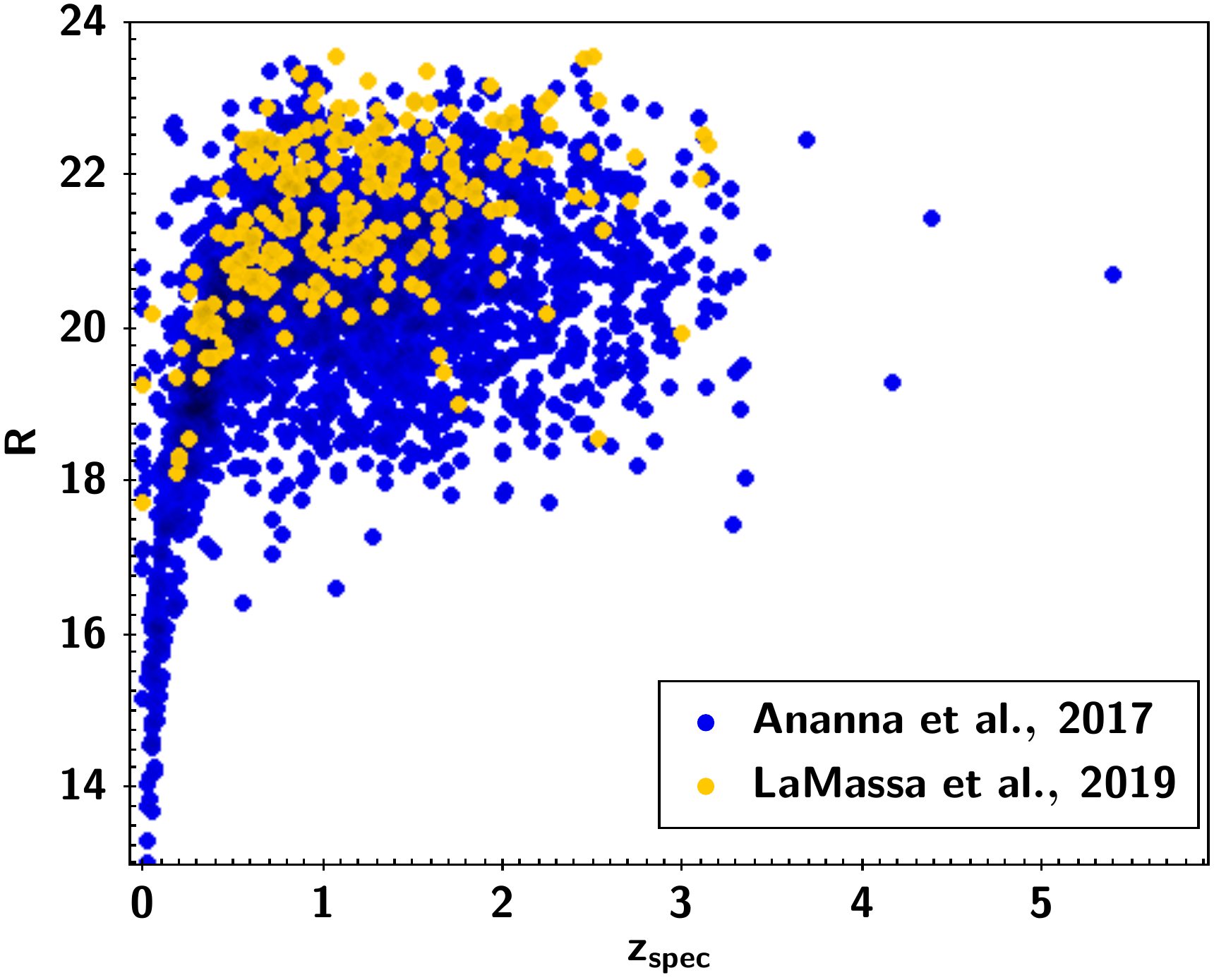}
\caption{Redshift and magnitude distribution for the sources with spectroscopic redshift. The blue sources were presented in \citetalias{Ananna2017} and have been used in this work as training and blind test samples. The $258$ yellow sources are on average fainter and were recently presented in \protect\cite{LaMassa19}. They are used as additional blind test sample.}
\label{fig:zs_mag}
\end{figure}

The spectroscopic coverage of the field (see Fig.~\ref{fig:zs_mag}) is ideal for assessing the performances of photo-z for X-ray selected sources via ML. The spectroscopic surveys BOSS \citep{Dawson13} and eBOSS \citep{Delubac17} in the life time of SDSS, provide reliable redshifts for about 50\% of the sources ($2,962/5,990$).
In addition Stripe 82X was also suitable for a dedicated spectroscopic program during SDSS-IV, targeting specifically the counterparts to X-ray sources \citep[]{LaMassa19}. There, the exposure time was of at least two hours long, allowing the determination of the redshifts also for faint sources.
The training and testing samples are formed only by sources with redshift available at the time of the publication of \citetalias{Ananna2017}. However, as additional blind test, we checked the accuracy of the photo-z also using this new available spectroscopic sample of $257$ sources (see Sec.~\ref{sec:pointEst}).
%%%%%%%%%%%%%%%%%%%%%%%%%%%%%%%%%%%%%%%%%%%%%%%%%
%%%%%%%%%%%%%%%%%%%%%%%%%%%%%%%%%%%%%%%%%%%%%%%%%%%%%%%%%%%%%%

\section{The Algorithms}\label{sec:Method} 

We first performed a feature analysis on the sub-sample of sources located in the yellow area of Fig.~\ref{fig:radec}. This sample maximizes the number of sources, the number of photometric bands, their depth and the faintest magnitudes available that are reported in the column $BEST$ of Table~\ref{tab:depth}.
The feature analysis  was performed  with the algorithm $\Phi$LAB, developed by our group and described in Sec.~\ref{sec:philab}.
Obviously the computation of the photo-z using the best features would provide the best accuracy, but it would dramatically limit the size of the sample for which the photo-z could be computed. 
In fact, because in this work magnitudes were used instead of fluxes, the sample includes many source missing data in various bands\footnote{working with fluxes would have allowed to use also the faint sources at background level, with negative fluxes but associated to a large, positive photometric errors. By definition these sources are not present in a catalogue that lists magnitudes.}. Rather than using specific but not yet fully tested methods for recovering the missing data \cite[see][for a discussion]{Brescia18}, we  computed the photo-z for subsets of sources that share the same photometric system (i.e., sources for which a certain combination of bands does not include missing data). The sub-samples naming convention is presented in Table~\ref{tab:depth}, together with the breakdown of the sources and the depth reached in every filter. The photo-z were estimated with MLPQNA (described in Sec.~\ref{sec:mlpqna}).
For each source we also derived the redshift PDZ using METAPHOR, described in Sec.~\ref{SEC:METAPHOR}.

The three algorithms are described in the following subsections.
%%%%%%%%%%%%%%%%%%%%%%%%%%%%%%%%%%%%%%%%%%%%%
\subsection{\mathinhead{\Phi}{Phi}LAB}\label{sec:philab}
Recently, in \cite{delliveneri19}, we presented a novel method suitable for a deep analysis and optimization of any Parameter Space, which provides in output the selection of the most relevant features. The algorithm developed by our group is called $\Phi$LAB (PhiLAB, Parameter handling investigation LABoratory). 
It is a hybrid approach, including properties of both wrappers and embedded feature selection categories \citep{Tangaro:2015}, based on two joined concepts, respectively: \textit{shadow features} \citep{Kursa10} and \textit{Na\"{\i}ve LASSO} (Least Absolute Shrinkage and Selection) statistics \citep{tibshirani12}. 
Shadow features are randomly noised versions of the real ones and their importance percentage is used as a threshold to identify the most relevant features among the real ones.
Afterwards, the two algorithms, based on LASSO and integrated into $\Phi$LAB, perform a regularisation, based on the standard $L_1$ norm, 
of a ridge regression on the residual set of weak relevant features (i.e. a shrinking of large regression coefficients to avoid overfitting). 
This has the net effect of \emph{sparsifying} the weights of the features, effectively turning off the least informative ones.

LASSO acts by conditioning the likelihood with a penalty on the entries of the covariance matrix and such penalty plays two important 
roles. First, it reduces the effective number of parameters and, second,  produces an estimate which is sparse.
Having a regularization technique as part of a regression minimization law, represents the most evident difference with respect to more 
traditional parameter space exploration methods, like PCA \citep{Pearson2010}. The latter is a technique based on feature covariance matrix 
decomposition, where the principal components are retained instead of the original features. The two concepts, \textit{shadow features} 
and \textit{Na\"{\i}ve LASSO}, are then combined within the proposed method by extracting the list of candidate most relevant features through 
the noise threshold imposed by the shadow features and by filtering the set of residual weak relevant features through the LASSO statistics.

$\Phi$LAB is detailed in \cite{delliveneri19}, where the method has been used to investigate the parameter space in the case of the photometric determination of star formation rates in the SDSS.

%%%%%%%%%%%%%%%%%%%%%%%%%%%%%%%%%%%%%%%%%%%%
\subsection{MLPQNA}\label{sec:mlpqna}
MLPQNA (Multi Layer Perceptron trained with Quasi Newton Algorithm) is a Multi Layer Perceptron (MLP; \citealt{Rosenblatt61}) neural network trained by a learning rule based on the Quasi Newton Algorithm (QNA) which is among the most used feed-forward neural networks in a large variety of scientific and social contexts, such as electricity price \citep{AGGARWAL200913}, detection of premature ventricular contractions \citep{EBRAHIMZADEH2010103}, forecasting stock exchange movements \citep{MOSTAFA20106302}, landslide susceptibility mapping \citep{Zare13} etc.\\
Furthermore, it has been successfully applied several times in the context of photometric redshifts (see for instance, \citealt{Biviano13,Brescia13,Brescia14a,Cavuoti15,de_Jong:17,Nicastro18}). The analytical description of the method has been discussed in the contexts of both classification \citep{Brescia20153893} and regression \citep{Cavuoti12, Brescia13}.

%%%%%%%%%%%%%%%%%%%%%%%%%%%%%%%%%%%%%%%%%%%%%%%%%%%%%%%%%%
\subsection{METAPHOR}\label{SEC:METAPHOR}

METAPHOR (Machine-learning Estimation Tool for Accurate PHOtometric Redshifts; \citealt{Cavuoti17}) is a modular workflow, designed to produce the redshift PDZs through ML. Its internal engine is the MLPQNA already described in Sec.~\ref{sec:mlpqna}.
The core of METAPHOR lies in a series of different perturbations of the photometry in order to explore the parameter space of data and to grab the uncertainty due to the photometric error.
In practice, the procedure to determine the PDZ of individual sources can be summarized in this way: we proceed by training the MLPQNA model and by perturbing the photometry of the given blind test set to obtain an arbitrary number $N$ of test sets with a variable photometric noise contamination. Then we submit the $N + 1$ test sets (i.e., $N$ perturbed sets plus the original one) to the trained model, thus obtaining $N + 1$ estimates of photo-z. With such $N + 1$ values we perform a binning in photo-z ($0.01$ for the described experiments), thus calculating for each one the probability that a given photo-z value belongs to each bin. In this work we used $N = 999$ to obtain a total of $1000$ photo-z estimates.

\subsection{Statistical Estimators}\label{SEC:statindicators}
For brevity, we define $\Delta z$ as:
\begin{equation}
\Delta z = (\zp-\zs)/(1+\zs)
\end{equation}

Then, in order to be able to compare the accuracy with that available for other surveys/ methods present in literature,  we use the classical basic statistical estimators, applied on $\Delta z$, described as following:\\
$\bullet$ mean (or bias);\\
$\bullet$ standard deviation $\sigma$;\\
$\bullet$ $\sigma_{NMAD} = 1.4826 \times median (|\Delta z|)$;\\
$\bullet$ $\sigma_{68}$ that is the width in which falls the $68$\% of the $\Delta z$ distribution;\\
$\bullet$ $\eta$, defined as the fraction (percentage) of outliers or source for which $|\Delta z| > 0.15$.\\

Due to the limited number of data samples, a canonical splitting of the dataset (or knowledge base) into training and blind test set cannot be applied. Therefore, in order to circumvent this problem the training+test process involves a k-fold cross validation \citep{hastie09,Kohavi95astudy}: the knowledge base has been manually split into $4$ not-overlapped sub-sets. In this way by taking each time $3$ of these sub-sets as training set and leaving the fourth as blind test set, an overall blind test on the entire knowledge base sample can be performed, i.e., each object of the available data sample has been evaluated in a blind way (i.e., not used for the training phase).

\section{Results on parameter space optimization}\label{sec:ps}
We started our experiments by considering the $BEST$ sample that maximizes the number of sources with spectroscopic redshift and with the maximum number of photometric points available (as discussed in Sec.~\ref{sec:photo}). The $BEST$ sample includes fifteen bands:
\begin{itemize}
\item u, g, r, i, z;
\item  J, H, K;
\item  CH1, CH2;
\item  W1, W2, W3, W4;
\item X-FLUX.
\end{itemize}

\begin{figure}
\centering
	\includegraphics[width=\columnwidth,trim={0 0.5cm  0 0.5cm},clip]{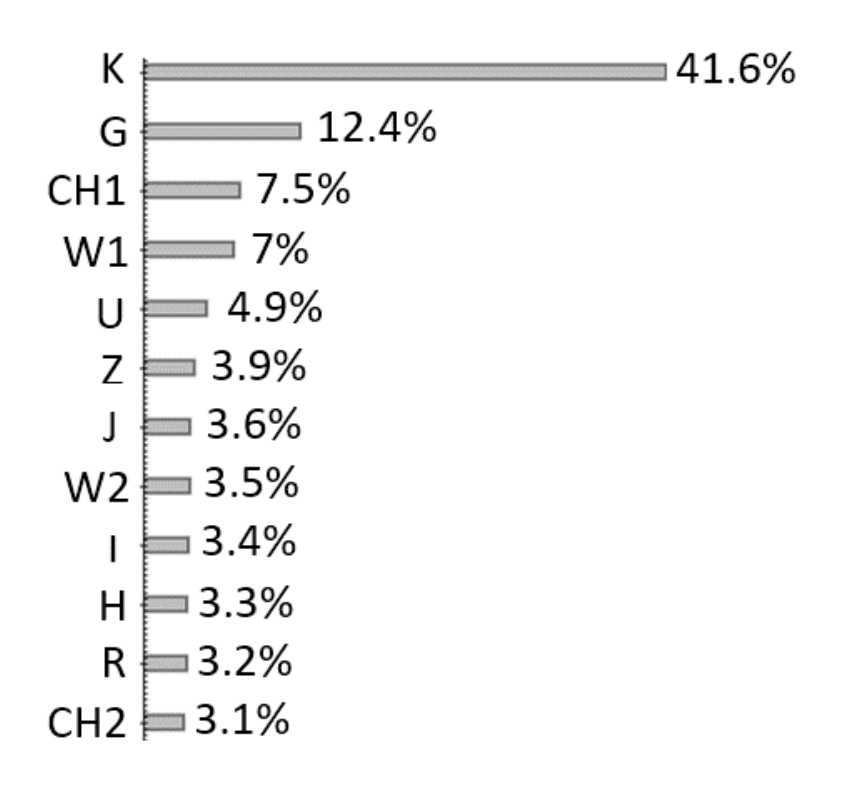}
\caption{Results of the feature analysis performed with $\Phi$LAB. The importance of each feature is estimated for the case in which only magnitudes are considered for the sample  $BESTmagopt$.}
\label{fig:fsmaghisto}
\end{figure}

\begin{figure}
\centering
	\includegraphics[width=\columnwidth,trim={0 0.75cm  0 0.55cm},clip]{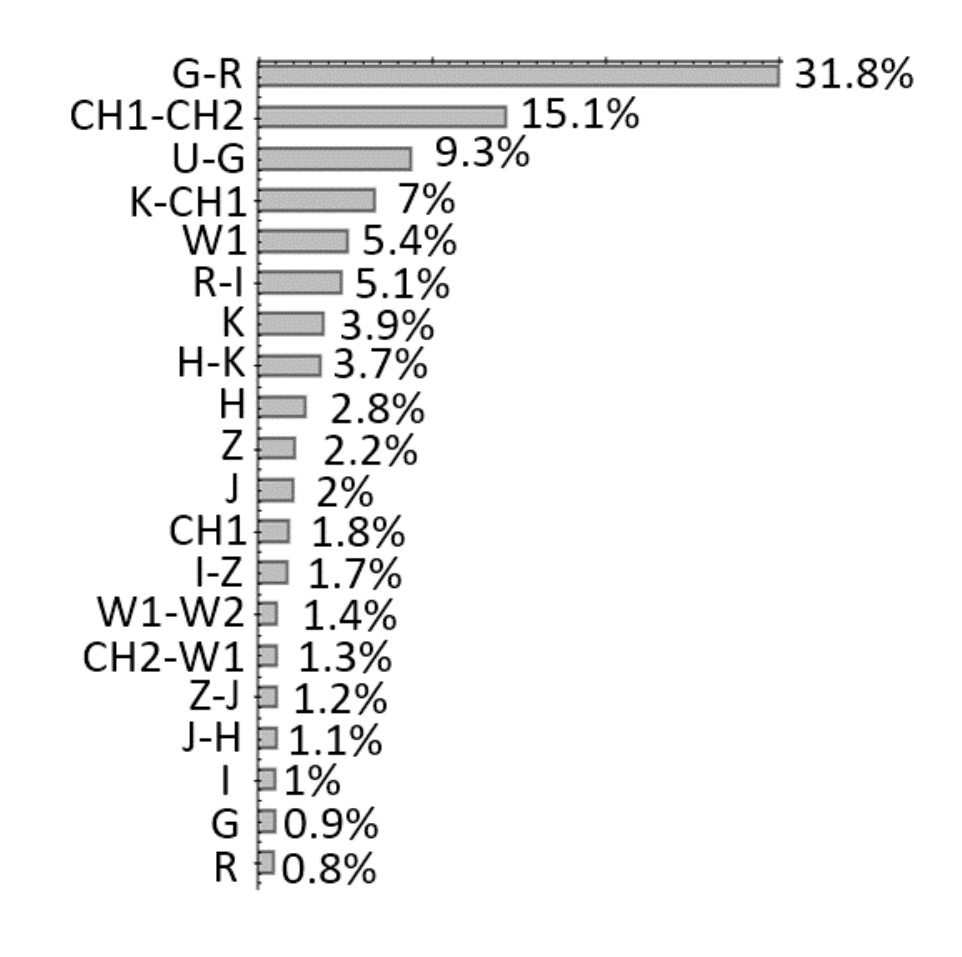}
\caption{Results of the feature analysis performed with $\Phi$LAB. The importance of each feature is estimated for the case in which magnitudes and colours are considered for the sample \textit{BESTmagcolopt}.}
\label{fig:fsmagcolhisto}
\end{figure}

Using $\Phi$LAB on the $BEST$ sample, we valuated the most relevant features, considering the parameter space of photometry alone, named $BESTmagopt$ (Fig.~\ref{fig:fsmaghisto}).
We then repeated the feature analysis on the $BESTmagopt$ space, in which we added the direct derived colours (i.e., those obtained by couples of adjacent magnitudes only), obtaining the optimized parameter space named as $BESTmagcolopt$. Fig.~\ref{fig:fsmagcolhisto} shows the impact of each selected feature in this case.
Finally, (as done in \citealt{Ruiz18}), we considered also the case of the full feature space, by including all $BESTmagopt$ magnitudes and all possible derived colours, for a total of $78$ features:

\begin{itemize}
\item all $5$ SDSS magnitudes and related $10$ colours;
\item all $3$ VHS magnitudes and related $3$ colours; 
\item all $2$ IRAC magnitudes and related one colour;
\item the two previously selected WISE bands W1, W2 and related one colour;
\item all $15$ combinations of colours among SDSS and VHS;
\item all $10$ combinations of colours among SDSS and IRAC;
\item all $10$ combinations of colours among SDSS and the two selected WISE bands;
\item all $6$ combinations of colours among VHS and IRAC;
\item all $6$ combinations of colours among VHS and the two selected WISE bands;
\item all $4$ combinations of colours among IRAC and the two selected WISE bands.
\end{itemize}

In this case, our method extracted a set of $67$ features considered suitable for the photo-z estimation. Table~\ref{tab:featureImp} reports the feature importance ranking for all the selected features. 
%%%%%%%%%%%%%%%%%%%%%%%%%%%%%%%%%%%%%%%%%%%%%%%
\begin{table}\centering
\begin{tabular}{lrlr}
\hline
  feature & importance & feature & importance\\
\hline
  R-Z & 14.51\% & J-K & 0.40\%\\
  G-I & 12.44\% & U-CH1 & 0.35\%\\
  CH1-CH2 & 7.50\% & H-CH1 & 0.34\%\\
  U-G & 6.00\% & R-CH2 & 0.33\%\\
  Z-W1 & 5.84\% & U-I & 0.33\%\\
  Z-CH1 & 4.24\% & R-W2 & 0.33\%\\
  G-R & 4.03\% & K-CH1 & 0.33\%\\
  K & 3.14\% & R-W1 & 0.31\%\\
  G-Z & 3.03\% & U & 0.30\%\\
  I-W1 & 2.00\% & U-J & 0.30\%\\
  I-CH2 & 1.94\% & G-W2 & 0.27\%\\
  H & 1.81\% & G-CH2 & 0.24\%\\
  R-I & 1.67\% & I-J & 0.23\%\\
  I-CH1 & 1.51\% & CH1-W2 & 0.23\%\\
  J & 1.45\% & G & 0.22\%\\
  H-K & 1.34\% & J-CH2 & 0.21\%\\
  R & 1.21\% & G-CH1 & 0.21\%\\
  I & 1.21\% & G-K & 0.20\%\\
  W1 & 1.18\% & J-W1 & 0.20\%\\
  I-Z & 1.08\% & H-W2 & 0.20\%\\
  Z & 0.99\% & K-CH2 & 0.17\%\\
  H-W1 & 0.97\% & K-W2 & 0.16\%\\
  K-W1 & 0.83\% & U-W1 & 0.16\%\\
  Z-W2 & 0.83\% & Z-J & 0.16\%\\
  CH2-W1 & 0.77\% & U-K & 0.15\%\\
  Z-CH2 & 0.68\% & R-CH1 & 0.14\%\\
  U-R & 0.68\% & H-CH2 & 0.13\%\\
  U-Z & 0.62\% & CH1 & 0.13\%\\
  G-W1 & 0.56\% & Z-H & 0.13\%\\
  J-CH1 & 0.54\% & U-H & 0.12\%\\
  Z-K & 0.52\% & J-W2 & 0.09\%\\
  I-W2 & 0.50\% & I-K & 0.08\%\\
  J-H & 0.46\% & R-K & 0.07\%\\
  W1-W2 & 0.45\% & --- &--- \\
\hline\end{tabular}
\caption{Results  of  the  feature  analysis  (percentages  of  estimated feature importance) performed with $\Phi$LAB in the case of the parameter space composed by considering all magnitudes and colours available.}\label{tab:featureImp}
\end{table}
%%%%%%%%%%%%%%%%%%%%%%%%%%%%%%%%%%%%%%%%%%%%%%%%%%%%%%%%%%%

Table~\ref{tab:preliminaryResults} reports the statistical results about the photo-z predictions for the various selected feature spaces. The photo-z estimation experiment performed with the largest selected parameter space (i.e., using the $67$ features selected by $\Phi$LAB among the $78$ available), provided statistical results comparable with the $BESTmagcolopt$ case, thus inducing to consider the latter as the best candidate parameter space, due to its smaller number of dimensions.
 When only magnitudes are considered, the K band is by far the most important feature. The reason is easily understood keeping in mind the SED of a galaxy. The rest frame K band indicates the knee of the SED and this clear feature is indeed suitable to determine the redshift.
However, the optimal feature combination turned out to be the mixed one, i.e., including colours and some reference magnitudes belonging to different surveys (for instance, \textit{GRIZ} for SDSS, \textit{JHK} for VHS, \textit{CH1} for IRAC and \textit{W1} for WISE). From the data mining viewpoint, this could appear rather surprising, since the kind of information should not be expected to change by introducing linear combinations between parameters, as colours are obtained by subtraction between magnitudes. However, from an astrophysical point of view, colours define the SED of the sources and their evolution with redshift, and for this reason they are crucial in the process of determining photo-z.
%However, as also discussed in \cite{Brescia13}, in more physical terms, this can be easily understood by considering that even though colours are obtained as a subtraction of magnitudes, the information content is different. Colours, in fact, implicitly introduce an ordering relationship within the parameter space, thus increasing the amount of information in the data (gradients instead of fluxes). 
The additional reference magnitudes, on the other hand, contribute to minimize the degeneracy in the luminosity class for a specific object type.\\
By keeping in mind such arguments, it appears less surprising that the information importance carried by magnitudes within the parameter space devoid of colours, drastically decreases in the mixed parameter space, where some representative colours result as the most significant features. In fact, the first four colour features of the $BESTmagcolopt$ parameter space (Fig.~\ref{fig:fsmagcolhisto}), contain about $65\%$ of the importance carried by the whole set of magnitudes present in the $BESTmagopt$ parameter space (Fig.~\ref{fig:fsmaghisto}). Furthermore, the strong relevance carried by the colours in the ranking list is reflecting the importance decreasing of related magnitudes, now represented by their colour combinations, causing in some cases the rejection of some magnitudes (e.g. U, CH2 and W2) from the optimized parameter space.

%-------------------------------------------------
%-------------------------------------------------

\subsection{Impact of feature analysis on photo-z}

The identification and consequent rejection of the non relevant features, allow us to obtain more accurate photo-z. 
This is demonstrated in Table~\ref{tab:preliminaryResults} where we report the accuracy and fraction of outliers for the photo-z computed with MLPQNA for the $BEST$ sample, with and without the removal of unimportant features for both $BESTmagopt$ and $BESTmagcolopt$. Table~\ref{tab:preliminaryResultsXray} is the same as Table~\ref{tab:preliminaryResults}, but this time the metrics are computed by limiting the samples to the sources that eROSITA will be able to detect. The comparison between the two tables points out something that is already well documented in literature in the case of photo-z computed via SED-fitting: namely, the accuracy of photo-z for AGN increases when the sample includes faint AGN, dominated by the host galaxies, easier to be modeled.
As soon as the sample is limited to bright AGN, the quality of photo-z decreases.

%The improvement is visible for each metric considered and the best result is obtained when the photo-z are computed considering not only photometric points but also some combination of colours. 

%-----------------------------------------
\begin{table}
\centering
\begin{tabular}{ccccc}
ID				&	$BEST$	&	$BESTmagopt$& $BESTmagcolopt$\\\hline
N. of Sources   &   1686        & 1686      &1686\\
bands			&	14		&	12 			& 20\\
|bias|			&	0.0159	&	0.0105		&0.0102\\
$\sigma$		&	0.141	&	0.135		&0.121\\
$\sigma_{NMAD}$	&	0.079	&	0.074		&0.056\\
$\sigma_{68}$	&	0.091	&	0.083		&0.069\\
$\eta$	        &	16.09	&	13.88		&12.74\\\hline
\end{tabular}
\caption{Accuracy of photo-z computed with MLPQNA on \textit{BEST}, \textit{BESTmagopt} and \textit{BESTmagcolopt} samples, after the optimization of the parameter spaces with the features analysis and selection performed with $\Phi$LAB. All quantities are calculated on blind test sets only.} \label{tab:preliminaryResults} %tabelle 4 e 6 riunite

\end{table}
%-----------------------------------------
\begin{table}
\centering
\begin{tabular}{ccccc}
ID				&	$BEST$	&	$BESTmagopt$& $BESTmagcolopt$   \\\hline
N. of Sources   &   1029    &   1029        &   1029            \\
bands			&	14      &	12			&   20               \\
|bias|			&	0.0157  &   0.0183	    &   0.0130           \\
$\sigma$		&	0.144   &	0.138       &   0.122            \\
$\sigma_{NMAD}$	&	0.078   &	0.072	    &   0.057            \\
$\sigma_{68}$	&	0.092   &	0.077	    &   0.074            \\
$\eta$	        &	15.90   &	12.31	    &   13.12            \\\hline
\end{tabular}
\caption{Same of Table~\ref{tab:preliminaryResults}, but considering only objects with F$_X>10^{-14}$.} \label{tab:preliminaryResultsXray}
\end{table}
%-----------------------------------------

\section{The impact of X-ray flux, photometry and morphology in the quality of photo-z}\label{sec:tomo}
After having identified the most relevant features, we were interested in exploring under which range of parameters the photo-z could be further improved. We have investigated in this respect the impact that photometric errors, X-ray depth and lack of information on optical morphology have on the results, again using the sample $BESTmagcolopt$ as reference.

It is worth noting that ideally, we should have also checked whether the feature analysis itself was influenced by these parameters. However, the samples are not sufficiently large for such kind of test. Therefore, we assume that the feature analysis is not affected and simply quantify the accuracy in photo-z for different sub-samples extracted from $BESTmagcolopt$, using different selection criteria reported in the following sections.

\subsection{Impact of Photometric Errors}\label{sec:magerr}
The photometric catalogues used in Stripe 82X are relatively shallow, implying that the fainter sources in general have a large photometric error. Unlike SED-fitting, until recently ML techniques 
could not handle the errors associated to the measurements and the same weight was incorrectly assumed for each photometric value \citep[but see][for a counter example application]{Reis19}. We tried to assess how this is impacting on the result, by reducing $BESTmagcolopt$ to sub-samples of decreasing photometric errors in all the bands. More specifically, we considered only sources with an error in magnitude
smaller than $0.3$, $0.25$ and $0.2$, thus reducing the original sample by $\sim15$\%, $\sim19$\% and $\sim24$\%, respectively (see Table~\ref{tab:magcut}).

%%%%%%%%%%%%%%%%%%%%%%%%%%%%%%%%%%%%%%%%%
\begin{table}\resizebox{\columnwidth}{!}{
\centering
\begin{tabular}{ccccc}
& \multirow{2}{*}{$BESTmagcolopt$} & \multicolumn{3}{c}{with Mag\_err limit}\\
  				&  			& 0.3 		& 0.25 		& 0.2		\\\hline
N. of sources &    1686     &     1442      &     1372      &    1275       \\  		
|bias|  		& 0.0102	& 0.0152 	& 0.0126	& 0.0122	\\
$\sigma$		& 0.121 	& 0.134  	& 0.157		& 0.151		\\
$\sigma_{NMAD}$ & 0.056	    & 0.056  	& 0.059 	& 0.054		\\
$\sigma_{68}$  	& 0.069		& 0.065	 	& 0.069		& 0.065		\\
$\eta$  		& 12.74	    & 11.93 	& 12.90     & 12.24	    \\
\end{tabular}}

\caption{Statistical results of the \textit{BESTmagcolopt} photo-z estimation experiments after having removed objects with photometric errors larger than $0.3$, $0.25$ and $0.2$ respectively.}\label{tab:magcut}%tabelle 7-8-9
\end{table}
%%%%%%%%%%%%%%%%%%%%%%%%%%%%%%%%%%%%%%
Considering only sources with small photometric errors provides the best accuracy ($\sigma_{NMAD}$=$0.054$), but the bias increases with respect to the original sample. The best trade off is obtained by keeping sources with a photometric error smaller than $0.3$ magnitudes.

\subsection{X-ray depth}\label{sec:xdepth}

The general experiment on Stripe 82X has demonstrated that reliable photo-z of the same quality, or even better than those computed via SED fitting, can be obtained for X-ray detected AGN also with ML, as long as a large number of photometric points is available and the spectroscopic sample is representative. But we are also interested to evaluate the accuracy obtained for a sub-sample of the sources, such as the brightest or the faintest detected in X-ray. And, in particular, the expected accuracy that can be reached for eROSITA. The final depth after four years of observations will be of  $\sim 10^{-14}erg/s/cm^{-2}$ for the all-sky survey. The survey will detect AGN also at high redshift, but it will be dominated by bright, nearby AGN, for which the computation of the photo-z is typically more challenging.
Table~\ref{tab:preliminaryResultsXray} already reported a partial answer to the questions: namely, the accuracy in photo-z for X-ray bright AGN is worse than for the entire sample. This means that the good results obtained in the second column are driven by the fact that the faint AGN, easier to fit because galaxy dominated, are more numerous than the bright AGN. However, in that experiment the training in the photo-z computation was done using all the sources in the $BESTmagcolopt$, with the cut in X-ray flux done \textit{a posteriori} on the output.
In the following experiment instead, also the training sample is limited to the bright sources that eROSITA will detect. The result of test is presented in Table~\ref{tab:xray}. By comparing the last column of that table with the last column of Table~\ref{tab:preliminaryResultsXray}, we see that, while the bias and $\sigma_{\rm NMAD}$ remain unchanged, the fraction of outliers decreases. It means that we could improve our result, if, in addition to good photometry for the entire sample, we could increase the training sample of bright objects by the time when eROSITA survey will be available.
Photo-z computed via ML for X-ray selected sources in 3XMM-DR6 and 3XMM-DR7 where recently presented also in \citet{Ruiz18} and \citet{Meshcheryakov18}, respectively. While we are in overall agreement with the first, our results are less optimistic than those obtained by the second group. However, the results is not surprising when noting that their results are specifically obtained for QSO or type $1$ only, having as targets sources in ROSAT and 3XMM-DR7 that are presented in the spectroscopic catalog SDSS-DR14Q \citep{Paris18}. In our work, there is no any pre-selection and the sample includes QSO, type $1$ and type $2$ AGN and galaxies.
%%%%%%%%%%%%%%%%%%%%%%%%%%%%%%%%%%%%%%%%
\begin{table}
\centering
\begin{tabular}{ccc}
 					& $BESTmagcolopt$ &	 at eROSITA depth\\  \hline
N. of sources       &     1686     &    1029     \\ 
|bias|				&	0.010      &	0.013	\\
$\sigma$			&	0.121      &	0.142	\\
$\sigma_{NMAD}$		&	0.056      &	0.064	\\
$\sigma_{68}$		&	0.069      &	0.075	\\
$\eta$			    &	12.74      &	12.73	\\
\end{tabular}
\caption{Comparison between statistics for the complete best sample and for the sub-sample limited to eROSITA flux also in the training sample.\label{tab:xray}}
\end{table}
%%%%%%%%%%%%%%%%%%%%%%%%%%%%%%%%%%%%%%%%%%%

\subsection{Point-Like vs Extended}\label{sec:objtypes}

Given the resolution of the ground-based optical imaging, extended sources can only be found at low redshift (\zspec $\leq 1$), with not significant contribution to the emission by the host galaxy. In contrast, point-like sources are mostly dominating the high redshift regime, with the emission due to the nuclear component.
This is taken into account when computing photo-z via SED-fitting by adopting a prior in absolute magnitude \citep[e.g.,][ \citetalias{Ananna2017}]{Salvato09,Salvato11,Fotopoulou12,Hsu14}. More recently, the separation of the sources in these two subgroups is becoming the standard also when computing photo-z via ML \citep[e.g.,][]{Mountrichas17, Ruiz18}.
One limitation of this method is that it relies on images that are affected by the quality of the seeing, which can alter the morphological classification of the sources.
This has been demonstrated in \citet[]{Hsu14} where the authors shown how the sources can change classification (and thus their photo-z value), depending on whether the images used are from HST or ground based.
In Stripe 82X, out of $1469$ sources at z$>$1, $77$ ($\sim$5\%) are classified as extended. In the $BESTmagcolopt$ sample the fraction is approximately the same ($27/704$, $\sim$4\%). Given the resolution of SDSS, this is clearly nonphysical.\\ 
In this section we first measure separately the accuracy for the point-like and extended sources in the $BESTmagcolopt$ sample. Here, a mixed training sample was used. In our second approach we created two training samples: one that includes only sources classified as "extended" and having redshift smaller than one; a second including only sources classified as "point-like" and/or at redshift larger than one. The sources in the test samples were separated accordingly.\\
The resulting statistics are shown in Table~\ref{tab:objtype}, with the first column reporting for convenience the first one of the two previous tables.

The second and third columns show the accuracy for the same sources of the $BESTmagcolopt$ sample, but this time divided according to their extension.\\
As expected, the photo-z for extended sources are more reliable, with $50$\% less outliers than the point-like sources. Column four and five show the extreme case, in which the training sample is split in two \textit{ab initio}. In this case, the photo-z for extended sources have the same accuracy of the best photo-z for normal galaxies, virtually without outliers or bias.\\ 
In contrast, the photo-z for point-like sources show about $20$\% of outliers. This is clearly understood, again thinking about the SED of these objects, by comparing 
Col3 with Col5 of the table. This suggests that the training sample for point-like sources must include also sources with a contribution from the host. The last column of the table shows which precision for the entire sample is achieved by two training samples (one specific for extended and nearby sources and the second for a generic one).
%%%%%%%%%%%%%%%%%%%%%%%%%%%%%%%%%%%%%
\begin{table*}
\centering
\begin{tabular}{cccc||ccc}
                & $BESTmagcolopt$ & \multicolumn{2}{c||}{limited to:} & \multicolumn{3}{c}{with specific training:}\\
  				&   & Extended  & Point Like & Extended & Point Like   & Combined\\	
  				\hline
N. of sources   & 1686      &    598    &  1088  & 598      &  1088     & 1686   \\ 
|bias|  		& 0.0102	&  0.0097   & 0.0107 & 0.0005	& 0.0168    & 0.0099 \\
$\sigma$ 		& 0.121 	&   0.089   & 0.136  & 0.051	& 0.142     & 0.109	  \\
$\sigma_{NMAD}$ & 0.056		&   0.042   & 0.068  & 0.029	& 0.082     & 0.053	 \\
$\sigma_{68}$  	& 0.069		&  0.050    & 0.082  & 0.032	& 0.096     & 0.071	 \\
$\eta$  		& 12.74	    &  7.69     & 15.57  & 1.67   	& 19.40     & 12.59	 \\
\end{tabular}

\caption{Photo-z estimation accuracy for the sources in $BESTmagcolopt$ using a unique training sample (Col1), afterwards divided between Extended 
(Col2) and Point-Like (Col3). The photo-z are also computed by splitting the sources between the two groups and training them separately. 
For this case the accuracy for Extended and Point-Like are presented in Col4 and Col5. In Col6 we recombine the sample. The improvement can be 
seen by comparing the column \textit{Combined} with \textit{$BESTmagcolopt$}. All quantities are calculated on the blind test set extracted from the \textit{BESTmagcolopt} sample.} \label{tab:objtype} 
\end{table*}
%%%%%%%%%%%%%%%%%%%%%%%%%%%%%%%%%%%%%%%%%%%%%%%%

\section{Photo-z estimation and reliability}\label{sec:pointEst}
In the previous section we analysed the impact of different factors on the accuracy of photo-z for X-ray selected sources. However, that analysis was done only on a small sample of the sources in Stripe 82X, for which all the photometric points were available. It will be possible to use all what we have learned on tens of thousands of square degrees of sky only when surveys such as LSST \citep{Ivezic19} and SPHEREx \citep{Dore18} will come online.
For the time being, in case of photo-z estimated via ML using catalogues of photometric points expressed in magnitudes, we face the problem that for many sources some of the values are missing.
For this reason, in order to provide a photo-z for most of the sources, we prioritized the sample divided in subgroups that share the same multi-wavelength coverage, sorted by accuracy in terms of $\sigma_{\rm NMAD}$:
 
\begin{enumerate}
\item SDSS, VHS, WISE \& IRAC (\textit{sdssVWI})
\item SDSS, VHS \& WISE (\textit{sdssVW});
\item SDSS, VHS \& IRAC (\textit{sdssVI});
\item SDSS \& WISE (\textit{sdssW});
\item SDSS \& IRAC (\textit{sdssI});
\item SDSS \& VHS (\textit{sdssV});
\item SDSS.
\end{enumerate}

In addition, with the goal of comparing the results with those obtained via SED-fitting in \citetalias{Ananna2017}, we have created a sample, MLPQNA$_{\rm merged}$, where for each source we consider the photo-z  computed via MLPQNA using the dataset with the highest accuracy available.

\begin{figure*}
\centering
\includegraphics[width=\textwidth, trim={0cm 2.8cm 0 3cm},clip]{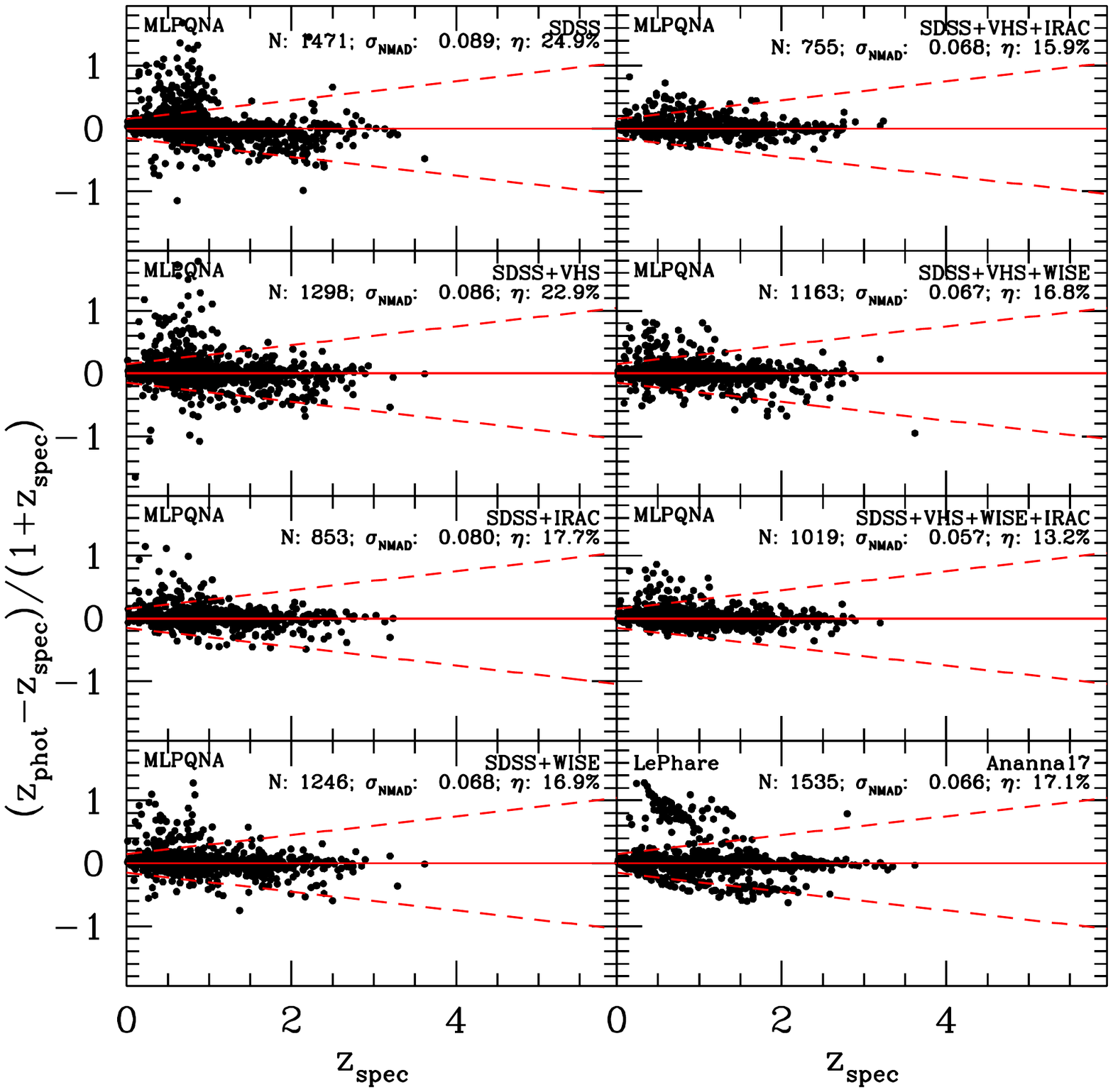}
\caption{Comparison between spectroscopic redshift and photo-z for the sources cut at the eROSITA flux and  divided on the basis of available photometric points. For comparison, the result from \citetalias{Ananna2017} is reported in the lower right panel of the figure. By comparing the accuracy and the fraction of outliers in every panel with the corresponding row in Table~\ref{tab:compl}, we see that computing photo-z using only SDSS for bright X-ray sources is not recommended.}
\label{fig:photo-z1}
\end{figure*}

This sorting, with the obvious limitation of producing photo-z with different accuracy across the field, offers nevertheless the possibility to characterize their quality as a function of the amount of photometric bands and the wavelength coverage.
The results of the metrics used for measuring the quality of the photo-z are presented in Table~\ref{tab:compl} for the entire sample 
and in Table~\ref{tab:commoncompl} for the sources that are in common to all samples. A visualisation of the results is also presented in Fig.~\ref{fig:photo-z1}, which  provides the comparison between spectroscopic and photometric redshifts computed with MLPQNA  for the sources in each sub-sample and in \citetalias{Ananna2017}. Here we plot only the sources that eROSITA will detect.
The photometric coverage, limited to all optical bands, reduces the sample by $5\%$, from 1535 down to 1471 and produces an excess of high redshift values for sources that are  actually at low redshift. The effect can be mitigated by adding redder bands, with the MIR bands from WISE being more efficient than the NIR bands from VHS.
The combined addition of NIR and MIR photometric points removes most of the outliers, but it also reduces the original sample by $35\%$ (from $1535$ to $1019$ sources). \\

\begin{figure*}
\centering
\includegraphics[width=\columnwidth, trim={0cm 0 0cm 0 cm},clip]{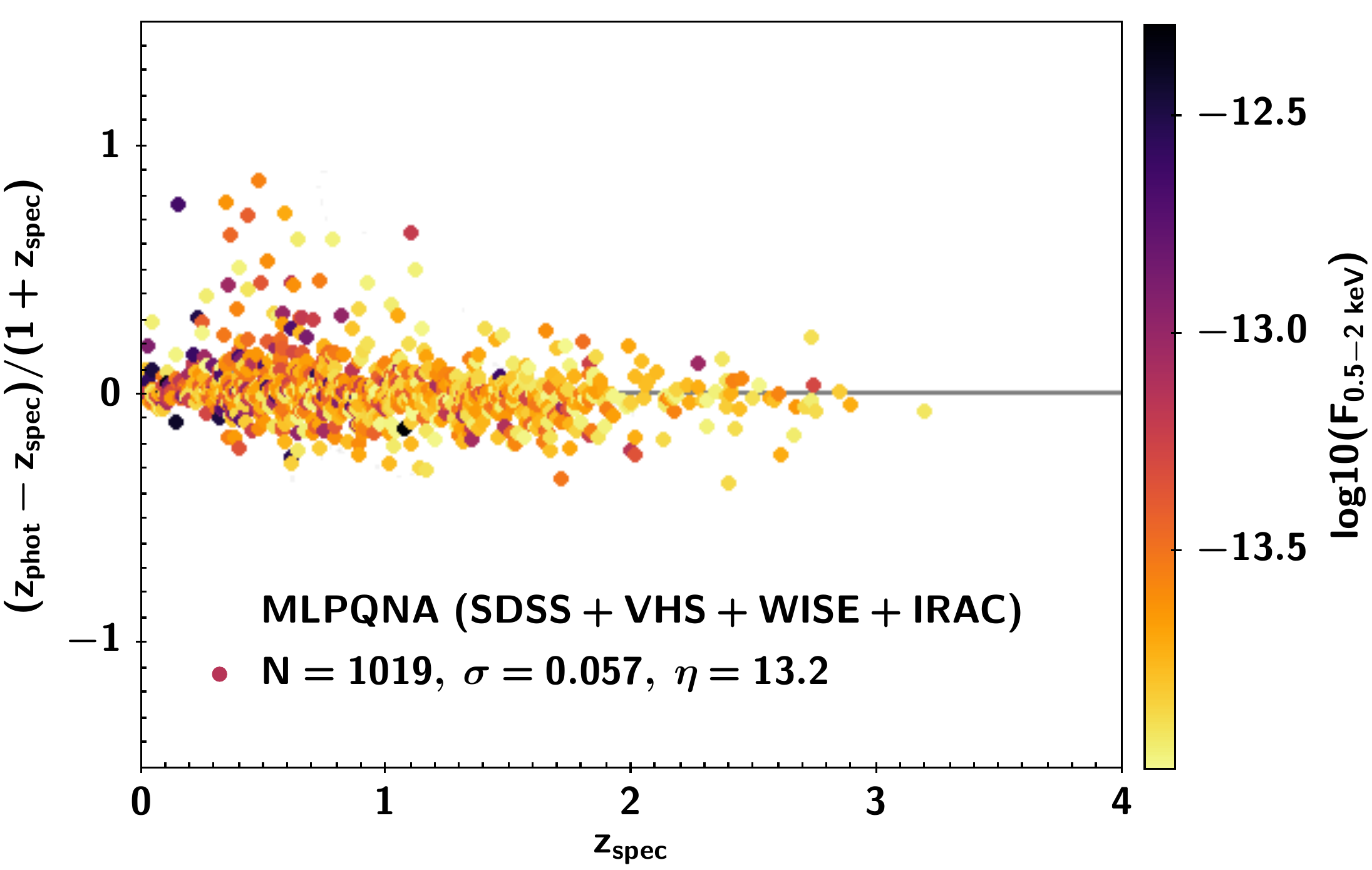}
\includegraphics[width=\columnwidth, trim={0cm 0 0cm 0 cm},clip]{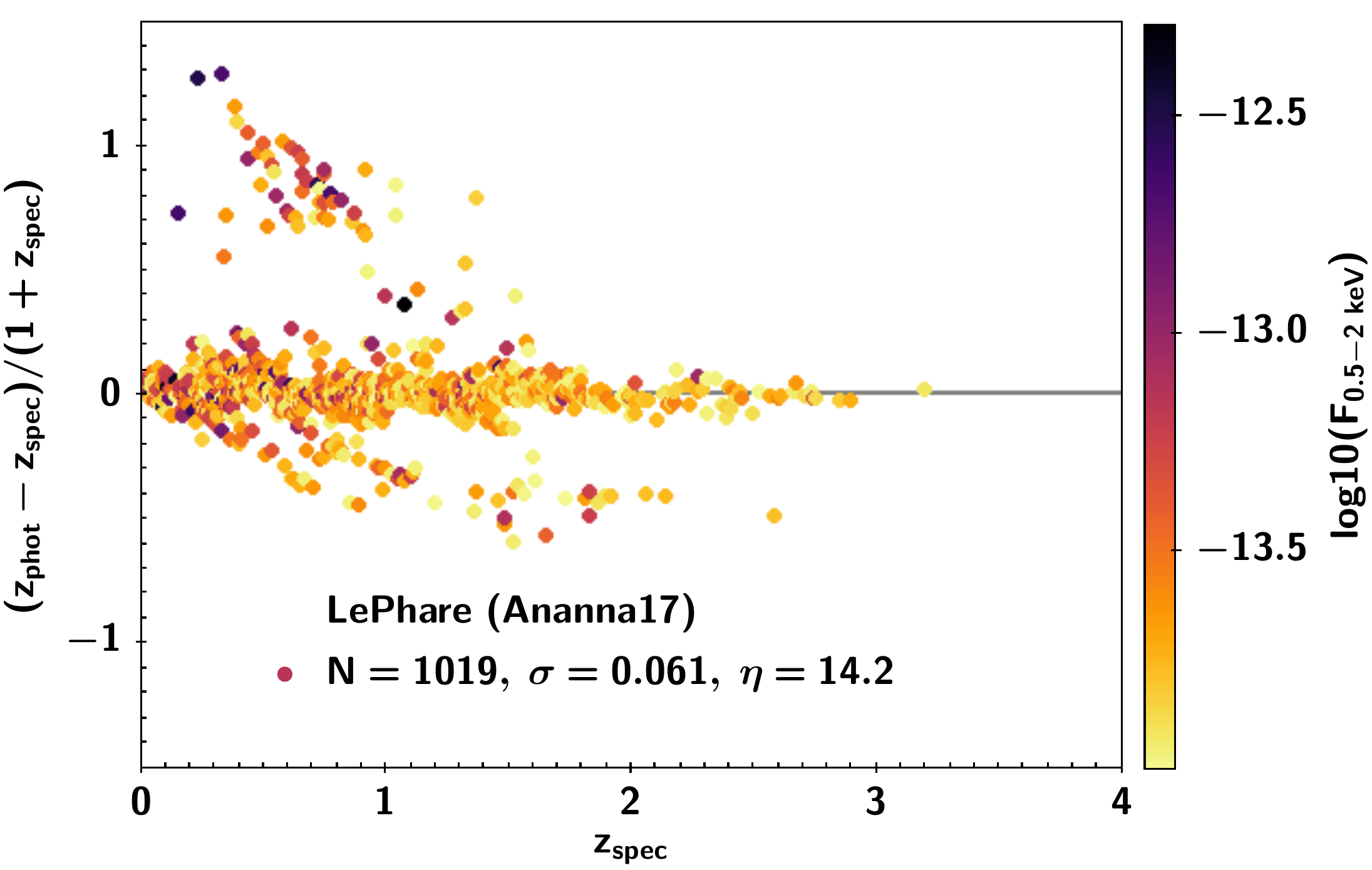}
\caption{The same as the last two bottom-right panels of Fig.~\ref{fig:photo-z1}, but this time limiting the sources from 
\citetalias{Ananna2017} to the same for which MLPQNA can compute photo-z using all the photometry available. The sources are colour coded 
as a function of their X-ray fluxes. With SED fitting, the outliers with overestimated redshift are basically X-ray bright.}
\label{fig:photo-z0}
\end{figure*}

\begin{figure}
\centering
\includegraphics[width=\columnwidth, trim={0cm 0 0cm 0 cm},clip]{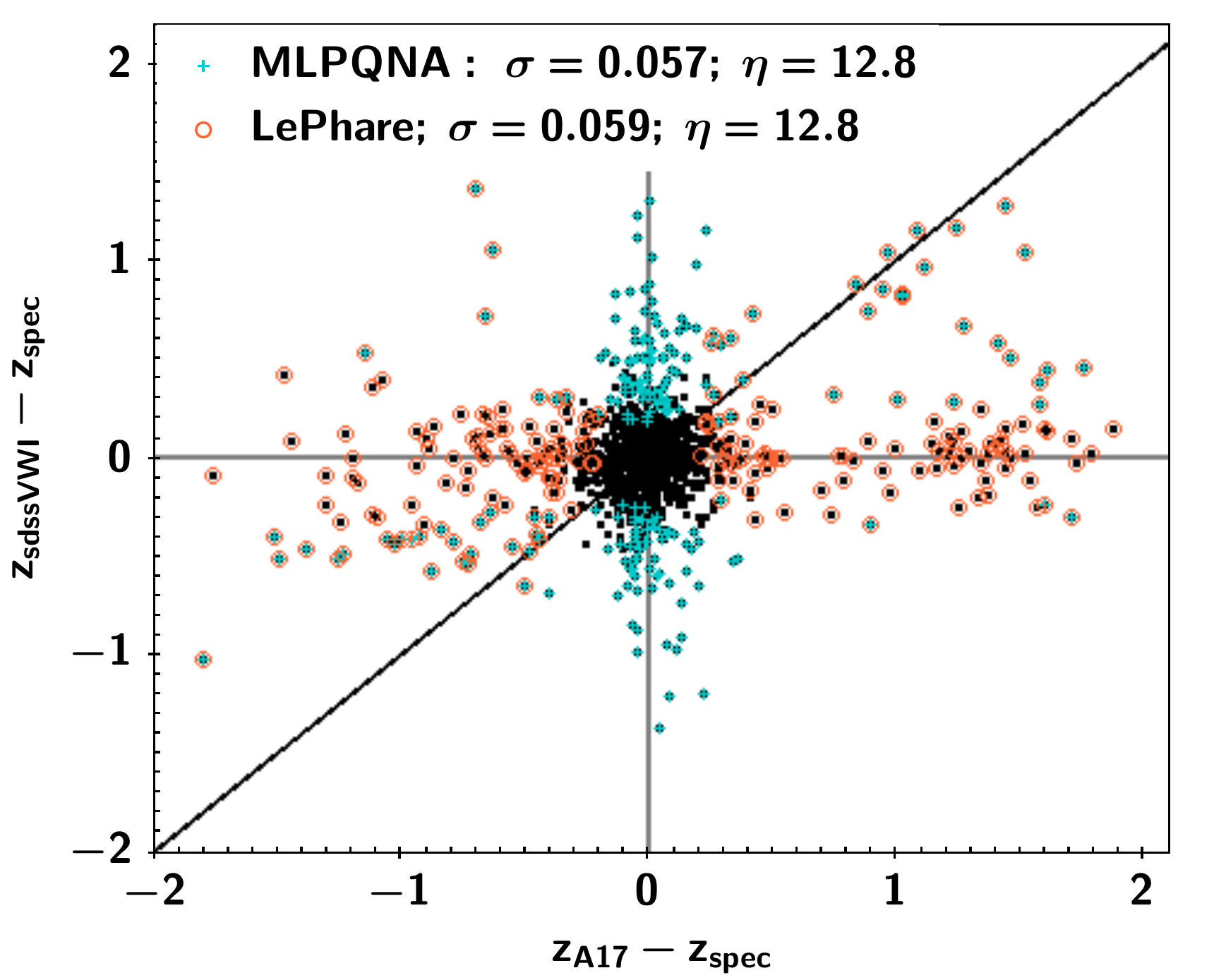}
\caption{Difference between spectroscopic redshift and photo-z computed via MLPQNA and LePhare for the sub-sample of 1689 sources with SDSS, VHS, WISE and IRAC photometry, regardless their X-ray flux. Sources that are outliers for MLPQNA (LePhare) are plot in cyan (orange). For this sub-sample, with complete photometry, the accuracy and fraction of outliers are very similar for the two methods. Nevertheless, the majority of the outliers are such only for one of the two algorithms.
For the common outliers along the  black one-to-one line, the two methods agree in terms of predicted photo-z.}
\label{fig:photo-z5}
\end{figure}

However, at the X-ray flux of eROSITA, and when SDSS, VHS, WISE and IRAC data are simultaneously available, MLPQNA performs better than the SED-fitting technique, with a lower fraction of  outliers and the reassuring absence of systematics (Fig.~\ref{fig:photo-z0}). The result is even more impressive if we consider the limited size of the training sample (contrary to \citealt[]{Ruiz18}, only spectroscopy available within the STRIPE82X field is used).\\
At the depth of eROSITA, the two methods are equally performing, with essentially the same accuracy and fraction of outliers (Fig.~\ref{fig:photo-z5}).

ML facilitates the process of computing reliable photo-z when the number of photometric points is sufficient, because it avoids any assumption on the type of templates needed in the library when SED-fitting technique is used. As underlined in \citetalias{Ananna2017}, this is a lengthy and risky procedure, since a slightly different set of template SEDs can  produce vastly different results.
However, when only a limited number of photometric points is available, SED-fitting remains a better approach, as it provides more reliable results and smaller fraction of outliers for the entire sample. 

Fig.~\ref{fig:photo-z1} also shows that the performance of sdssVWI is higher compared to sdssVW and sdssVI. This is due to the fact that, 
as can be seen in Fig. 4 of the official WISE web site\red{\footnote{\url{http://wise2.ipac.caltech.edu/docs/release/prelim/expsup/sec4_3g.html}}}, 
although W1 and W2 are centered almost at the same wavelength of IRAC/CH1 and IRAC/CH2, they are broader, with W1(W2) extending to 
shorter(longer) wavelength, with respect to CH1(CH2). Using simultaneously the four bands increases the characterisation of the SEDs.\\
IRAC photometry is deeper and more precise (i.e., smaller photometric errors) than WISE photometry  and this explains why, at the depth of 
eROSITA, sdssVI performs better than sdssVW. As we  discussed in subsection \ref{sec:magerr} a precise photometry helps in disentangling the 
correct SED of the sources and this is particularly true for the bright X-ray sources. In contrast, Table~\ref{tab:compl} shows 
that when the sample includes also faint sources, sdssVW performs better that sdssVI. This can be explained by considering that the sources 
start to be dominated by the host, with less differences in their SED and for this reason better determined using a larger wavelength coverage 
than a precise photometric point.\\
It is worth noting that, although IRAC data are available only for certain areas of the sky, the all-sky observations with WISE is ongoing. 
Already the most recent data release \citep{Schlafly19} is 0.7 magnitude deeper than the data used here.
So it is plausible to predict that the precision over the entire sky for eROSITA is well represented by the sdssVI case and will not be worse 
than the sdssVW case.
%%%%%%%%%%%%%%%%%%%%%%%%%%%%%%%%%%%%%%%%%%
\begin{table*}
\centering
\begin{tabular}{lcccccc||cc} 
	&	\multirow{2}{*}{\bf \shortstack{Number of\\ sources}}	&\multirow{2}{*}{\bf $|bias|$} & \multirow{2}{*}{\bf $\sigma$}	&		\multirow{2}{*}{\bf $\sigma_{68}$} & \multirow{2}{*}{\bf $\sigma_{\rm NMAD}$} & \multirow{2}{*}{\bf $\eta$} & \multirow{2}{*}{\bf $\sigma_{\rm NMAD_{\rm A17}}$}	& \multirow{2}{*}{\bf $\eta_{\rm A17}$}\\ 
	
	&&&&&&&\\
 \hline
  \hline
  
  (1) \textit{sdssVWI}    & 1686 & 0.0102 & 0.121 & 0.069 & 0.056 & 12.74 & 0.059 & 13.3\\
  (2) \textit{sdssVW}     & 1889 & 0.0163 & 0.169 & 0.077 & 0.065 & 16.20 & 0.060 & 13.2\\
 (3) \textit{sdssVI}      & 1279 & 0.0210 & 0.134 & 0.079 & 0.067 & 16.50 & 0.057 & 13.4\\
 (4) \textit{sdssW}	      & 2121 & 0.0193 & 0.190 & 0.078 & 0.069 & 16.55 & 0.061 & 13.8\\
 (5) \textit{sdssI}	      & 1595 & 0.0152 & 0.170 & 0.089 & 0.077 & 17.49 & 0.059 & 15.1\\
  (6) \textit{sdssV}	  & 2142 & 0.0247 & 0.264 & 0.108 & 0.089 & 23.24 & 0.060 & 13.9\\
   (7) \textit{sdss}      & 2747 & 0.0410 & 0.270 & 0.104 & 0.087 & 24.17 & 0.063 & 15.7\\
(8) MLPQNA$_{\rm merged}$ & 2780 & 0.0178 & 0.173 &0.082 &0.068 &16.51  & 0.063 & 15.9\\
\end{tabular}
\caption{Summary of all statistical results for the sub-samples with different photometric coverage, listed in decreasing order or reliability (based on $\sigma_{\rm NMAD}$). In MLPQNA$_{\rm merged}$ we list the best photo-z available for each source. All quantities are calculated on blind test sets. For comparison, on the last two columns we report $\sigma_{\rm NMAD}$ and $\eta$ for the same sub-sample in \citetalias{Ananna2017}. }\label{tab:compl}
\end{table*}
%%%%%%%%%%%%%%%%%%%%%%%%%%%%%%%%%%%%%%%%%%%%%%%%%%%%%%%%%%%%%

%%%%%%%%%%%%%%%%%%%%%%%%%%%%%%%%%%%%%%%%%%
\begin{table*}
\centering
\begin{tabular}{lccccc} 
	& {\bf $|bias|$} & {\bf $\sigma$}	&	{\bf $\sigma_{68}$} & {\bf $\sigma_{\rm NMAD}$} & {\bf $\eta$}\\ 
 \hline
  \hline
  
  (1) \textit{sdssVWI}  & 0.0075 & 0.119 & 0.070 & 0.059 & 11.99\\
  (2) \textit{sdssVW}   & 0.0147 & 0.166 & 0.077 & 0.065 & 16.57\\
  (3) \textit{sdssVI}   & 0.0164 & 0.126 & 0.078 & 0.065 & 15.07\\
  (4) \textit{sdssW}	& 0.0197 & 0.188 & 0.078 & 0.066 & 14.97\\
  (5) \textit{sdssI}	& 0.0096 & 0.143 & 0.088 & 0.075 & 15.07\\
  (6) \textit{sdssV}	& 0.0257 & 0.233 & 0.110 & 0.090 & 21.37\\
  (7) \textit{sdss}     & 0.0287 & 0.224 & 0.107 & 0.085 & 22.79\\
  (8) \textit{A17}      & - & - & - & 0.059 & 11.96\\
\end{tabular}
\caption{Same of Table~\ref{tab:compl}, but using only sources in common across all samples.}\label{tab:commoncompl}
\end{table*}
%%%%%%%%%%%%%%%%%%%%%%%%%%%%%%%%%%%%%%%%%%%%%%%%%%%%%%%%%%%%%

%%%%%%%%%%%%%%%%%%%%%%%%%%%%%%%%%%%%%%%%%%%%%%%%%%%%%%%%%%%%%

%%%%%%%%%%%%%%%%%%%%%%%%%%%%%%%%%%%%%%%%%%%%%%%%%%%%%%%
%%%%%%%%%%%%%%%%%%%%%%%%%%%%%%%%%%%%%%%%%%%%%%%%%%%%%%%

We usually assess the accuracy of photo-z using a single value and we tacitly assume that two independent methods with similar statistics will provide the same photo-z value for a given source. This is clearly not the case, as shown in Fig.~\ref{fig:photo-z5}, where, for the sources in \textit{sdssVWI} with spectroscopic redshift, we plot the difference between photo-z and zspec for the redshifts computed with MLPQNA (in cyan) and LePhare (in red).
It is interesting to note that only a small number of sources are simultaneously outliers in both methods. The majority are method-dependent, thus ruling out the possibility that these sources are outliers because they are peculiar objects (e.g., varying objects).

\begin{figure}
\centering
\includegraphics[width=\columnwidth,trim={0cm 0cm 0 0cm},clip]{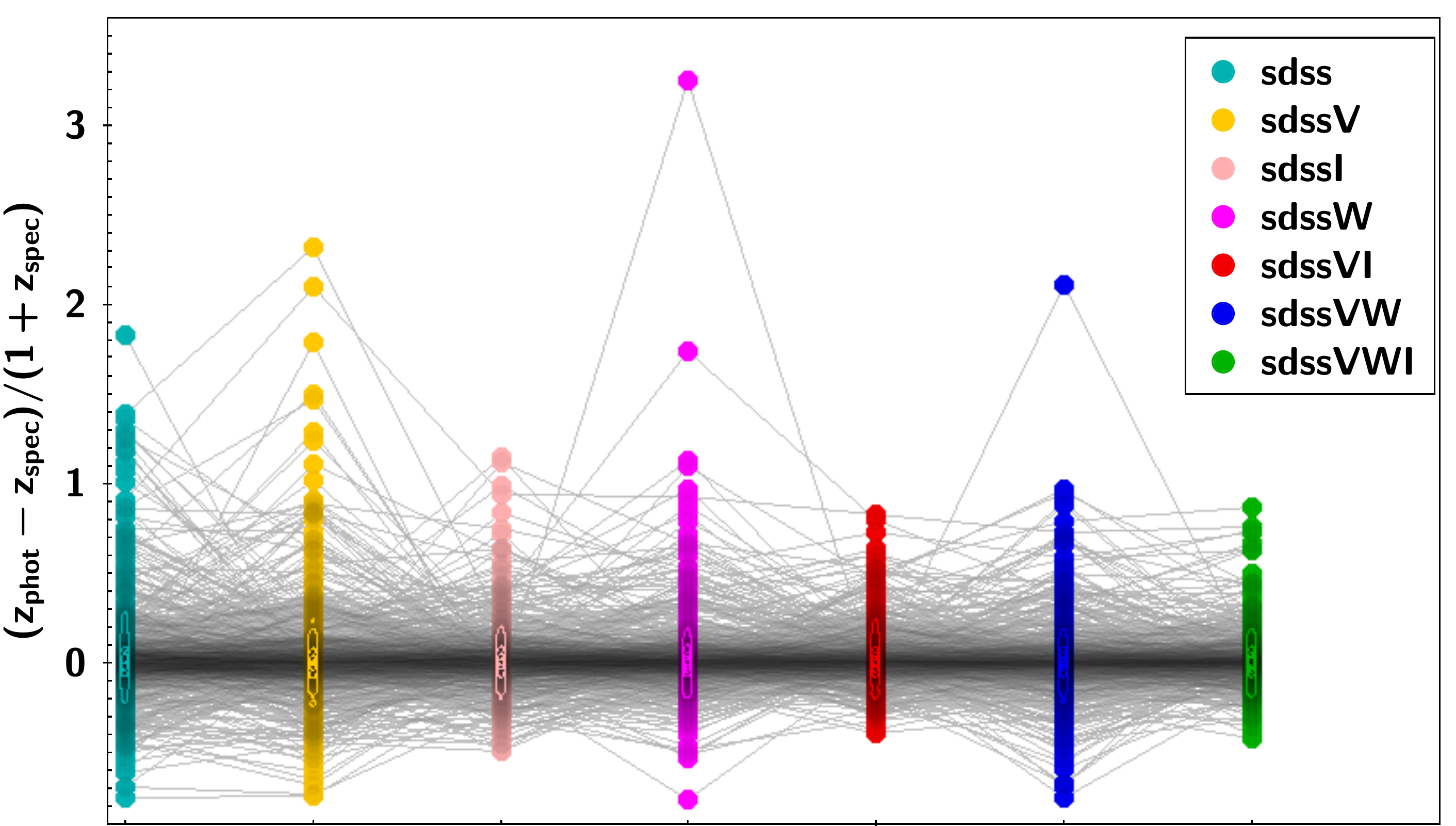}
\caption{One-to-one comparison of accuracy for photo-z computed via MLPQNA with different combinations of photometry. For this plot only sources present in all the subsamples have ben used.}
\label{fig:photo-z2}
\end{figure}

Fig.~\ref{fig:photo-z1} has already shown how, by adding more bands, the overall accuracy improves and the fraction of outliers decreases. 
But, are the outliers just being reduced or are there sources that became outliers with the increasing of the photometric bands? To this question 
we answer in Fig.~\ref{fig:photo-z2}. There, for each combination of bands, we see that there are sources having the correct photo-z computed 
with a limited number of bands and that become outliers when more bands are considered. These are a small fraction of the sources, and they can 
be explained by either unstable photometry (i.e., photometry with large errors), or by  a non representative spectroscopic sample.

%%%%%%%%%%%%%%%%%%%%%%%%%%%%%%%%%%%%%%%%%%%%%%%%%%%%%%%

\begin{figure*}
\centering
\includegraphics[width=\columnwidth,trim={0cm 1.5cm .5cm 2cm},clip]{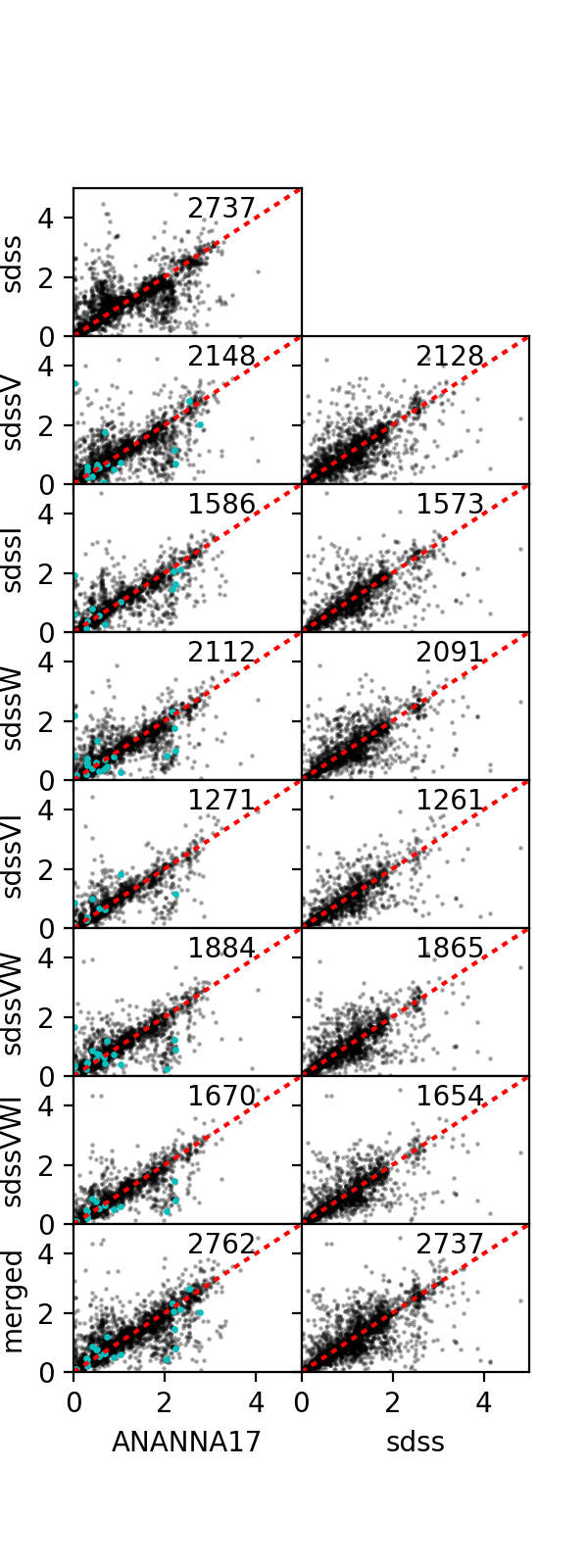}
\includegraphics[width=\columnwidth,trim={0cm 1.5cm .5cm 2cm},clip]{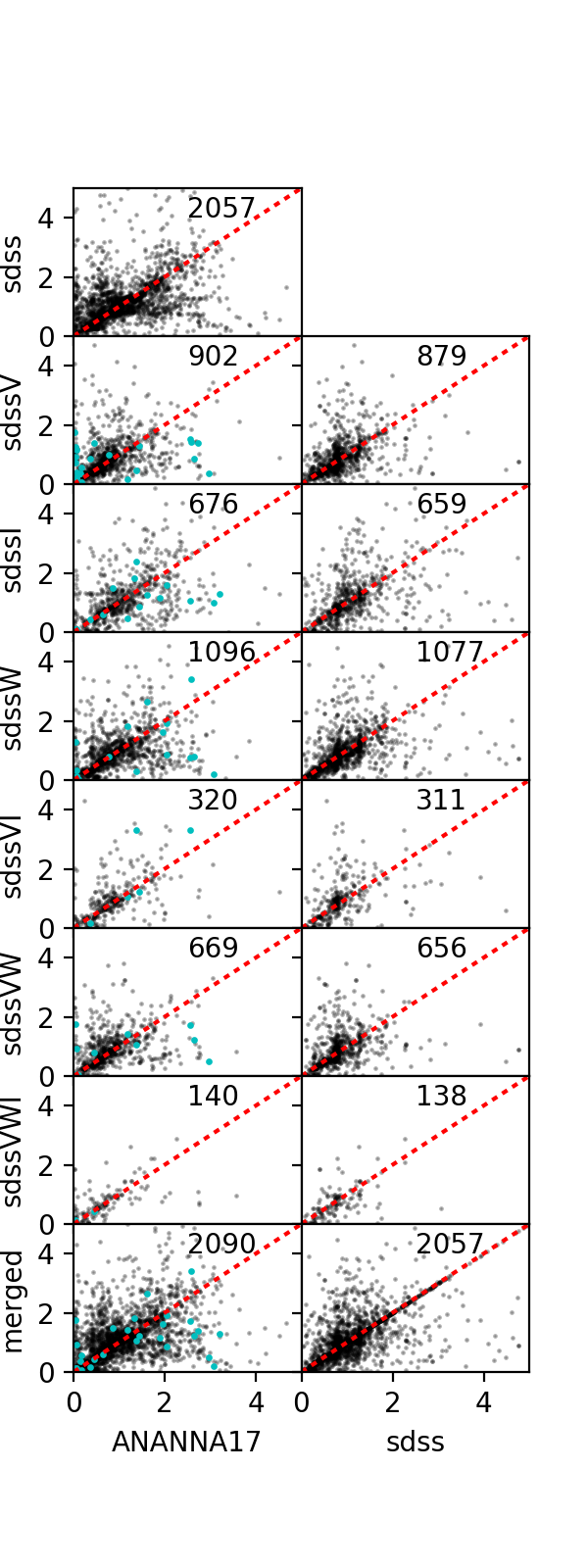}
\caption{Details of the comparison between photo-z computed via SED fitting (\citetalias{Ananna2017}) and MLPQNA for the sample for which spectroscopic information is, respectively, available (left panel) and not available (right panel). The cyan points indicate the sources for which the redshift could be computed only after considering supplementary photometry in addition to SDSS.}
\label{fig:photo-z32cols}
\end{figure*}

Finally, the left panel of Fig.~\ref{fig:photo-z32cols} shows, for each source with spectroscopic redshift, the details of the relative comparison between the photo-z computed via MLPQNA for the various photometric sets and LePhare from \citetalias{Ananna2017}.
Because all the photo-z incarnations used the same spectroscopic sample, the bulk of the sources in each plot lies on the one-to-one relation (red dotted lines). 
This is generally not the case when the same comparison is done using the sources for which no spectroscopic redshift is available (right panel of Fig.~\ref{fig:photo-z32cols}). 
 However, it is reassuring that, despite the small size of the sample, the photo-z computed with MLPQNA, using SDSSVWI sample, agree with those computed in \citetalias{Ananna2017}. In contrast, the second column of the right panel of Fig.~\ref{fig:photo-z32cols} shows how computing the photo-z using only SDSS creates an excess of sources at z $\sim 0.7$ and z $\sim 1.1$.  
The remaining comparison cases are reported, for completeness, in Appendix~\ref{sec:appendix} (Figures~\ref{fig:photo-z3allWzspec} and \ref{fig:photo-z3allNozspec}).

%%%%%%%%%%%%%%%%%%%%%%%%%%%%%%%%%%%%%%%%%%
%%%%%%%%%%%%%%%%%%%%%%%%%%%%%%%%%%%%
To further test our accuracy, we decided to compare the results against the $257$ new spectroscopic redshifts newly presented in \protect\citealt{LaMassa19}. The Table~\ref{tab:newz} shows the photo-z accuracy for MLPQNA and LePhare for this sub-sample of objects. Once again, the noticeable difference in the accuracy and fraction of outliers obtained (see Tables \ref{tab:compl} and \ref{tab:commoncompl} for the comparison) for every dataset, points to the importance of the training sample that must be representative of the entire population, in type, but also in fraction to the total. This new sample covers the same redshift range as the original one, but it is dominated by fainter sources (see Fig.~\ref{fig:zs_mag}). The same issue affects, although marginally, also the photo-z computed with SED-fitting. This is not surprising if one takes into account the fact that the templates, to be considered in the library, are determined by looking at the properties of the spectroscopic sample. However, in the case of SED-fitting the effect is minor, because only the SED type is considered and not its frequency in the sample.
%%%%%%%%%%%%%%%%%%%%%%%%%%%%%%%%%%%%%%%%%%
\begin{table*}
\centering
\begin{tabular}{lcccccc} 
	&	\multirow{2}{*}{\bf \shortstack{Number of \\ sources}}	&\multirow{2}{*}{\bf $|bias|$} & \multirow{2}{*}{\bf $\sigma$}	&		\multirow{2}{*}{\bf $\sigma_{68}$} & \multirow{2}{*}{\bf $\sigma_{\rm NMAD}$} & \multirow{2}{*}{\bf $\eta$} \\ \\
 \hline
  \hline
 A17     & 258 & 0.0066  & 0.292   & 0.129     & 0.089     & 27.07 \\
 sdss    & 227 & 0.0037  & 0.367   & 0.158     & 0.129     & 33.48 \\
 sdssV	 & 135 & 0.0357  & 0.322   & 0.211     & 0.149     & 41.48 \\
 sdssW	 & 144 & 0.0073  & 0.288   & 0.173     & 0.137     & 34.03 \\
 sdssI	 & 110 & 0.0119  & 0.202   & 0.184     & 0.163     & 40.91 \\
 sdssVW  & 111 & 0.0459  & 0.272   & 0.167     & 0.143     & 33.33 \\
 sdssVI	 &  58 & 0.0343  & 0.255   & 0.161     & 0.116     & 32.76 \\
 sdssVWI & 25  & 0.0298  & 0.151   & 0.152     & 0.104     & 32.00 \\
MLPQNA$_{\rm merged}$ 
         & 229 & 0.0182 & 0.270 & 0.192  & 0.154  &38.43  \\
\end{tabular}
\caption{Summary of all statistical results for the new sample of $258$ spectroscopic redshifts presented in \protect\citealt{LaMassa19}.}
\label{tab:newz}
\end{table*}
%%%%%%%%%%%%%%%%%%%%%%%%%%%%%%%%%%%
%\begin{table*}
%\resizebox{\columnwidth}{!}{
%\centering
%\begin{tabular}{cccccc} 
%	&	\multirow{2}{*}{\bf SDSS}	&\multirow{2}{*}{\bf\shortstack{ SDSS \& \\ VHS}}&	\multirow{2}{*}{\bf\shortstack{ SDSS \& \\ WISE}}	&	\multirow{2}{*}{\bf\shortstack{ SDSS, \\ VHS \& WISE}}	&	\multirow{2}{*}{\bf\shortstack{ SDSS, VHS,\\ WISE \& IRAC}}\\
%    & & & & \\ \hline
%data amount		&	2773    & 2199		&	2142	&	1917	&	1696	\\
%$|bias|$		&	0.0330	& 0.0247	&	0.0154	&	0.0113	&	0.0102\\ 
%$\sigma$		&	0.291	& 0.264		&	0.194	&	0.175	&	0.121\\ 
%$\sigma_{NMAD}$	&	0.089	& 0.089		&	0.069	&	0.066	&	0.056\\ 
%$\sigma_{68}$	&	0.108	& 0.108		&	0.080	&	0.078	&	0.069\\ 
%$\eta>0.15$		&	29.29\%	& 23.24\%	&	17.09\%	&	16.95\%	&	12.74\%\
%\end{tabular}
%}
%\caption{Summary of all statistical results of the experiments with a different photometric coverage. All quantities are calculated on blind test sets.}\label{tab:compl}
%\end{table*}

\section{PDZ estimation and reliability }\label{sec:pdzEst}
%%%%%%%%%%%%%%%%%%%%%%%%%%%%%%%%%%%%%%%%%
\begin{figure}
\centering
\includegraphics[width=\columnwidth, trim={0.98cm 0 1.1cm  1cm},clip]{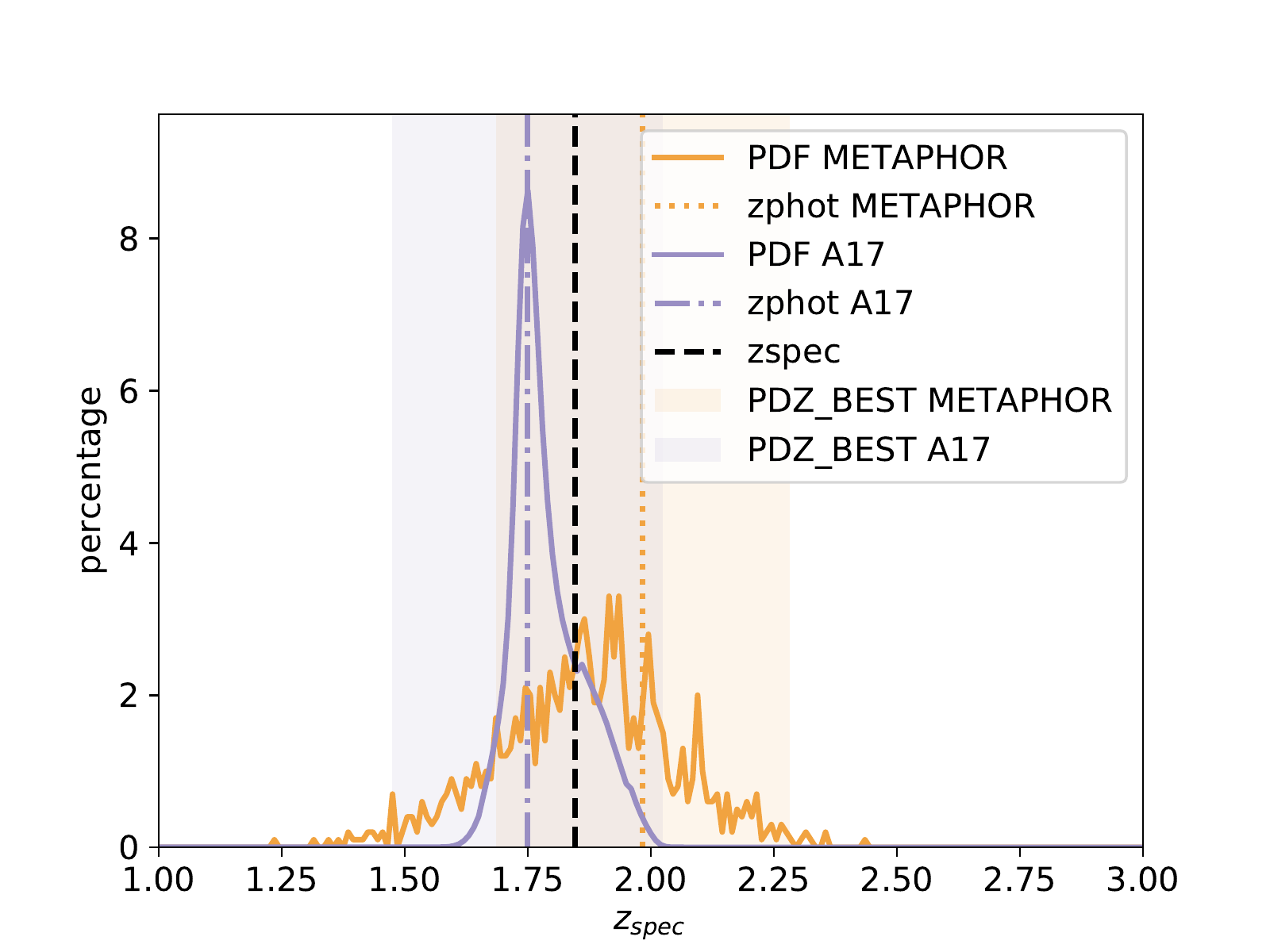}
\caption{Example of PDZ obtained by METAPHOR and LePhare  for the object \textit{ID1431} (\href{http://skyserver.sdss.org/dr14/en/tools/explore/summary.aspx?ra=333.7029&dec=0.41911}{SDSS J221448.69+002508.7}). The true redshift is represented by the black dashed line. The coloured areas represent the PDZ\_BEST. In this specific case, the PDZs are both limited to the redshift range 1-3 
for illustration purposes.}\label{fig:overlap}
\end{figure}
%%%%%%%%%%%%%%%%%%%%%%%%%%%%%%%%%%%%%%%%%
\begin{figure*}
\centering
\includegraphics[width=.49\textwidth]{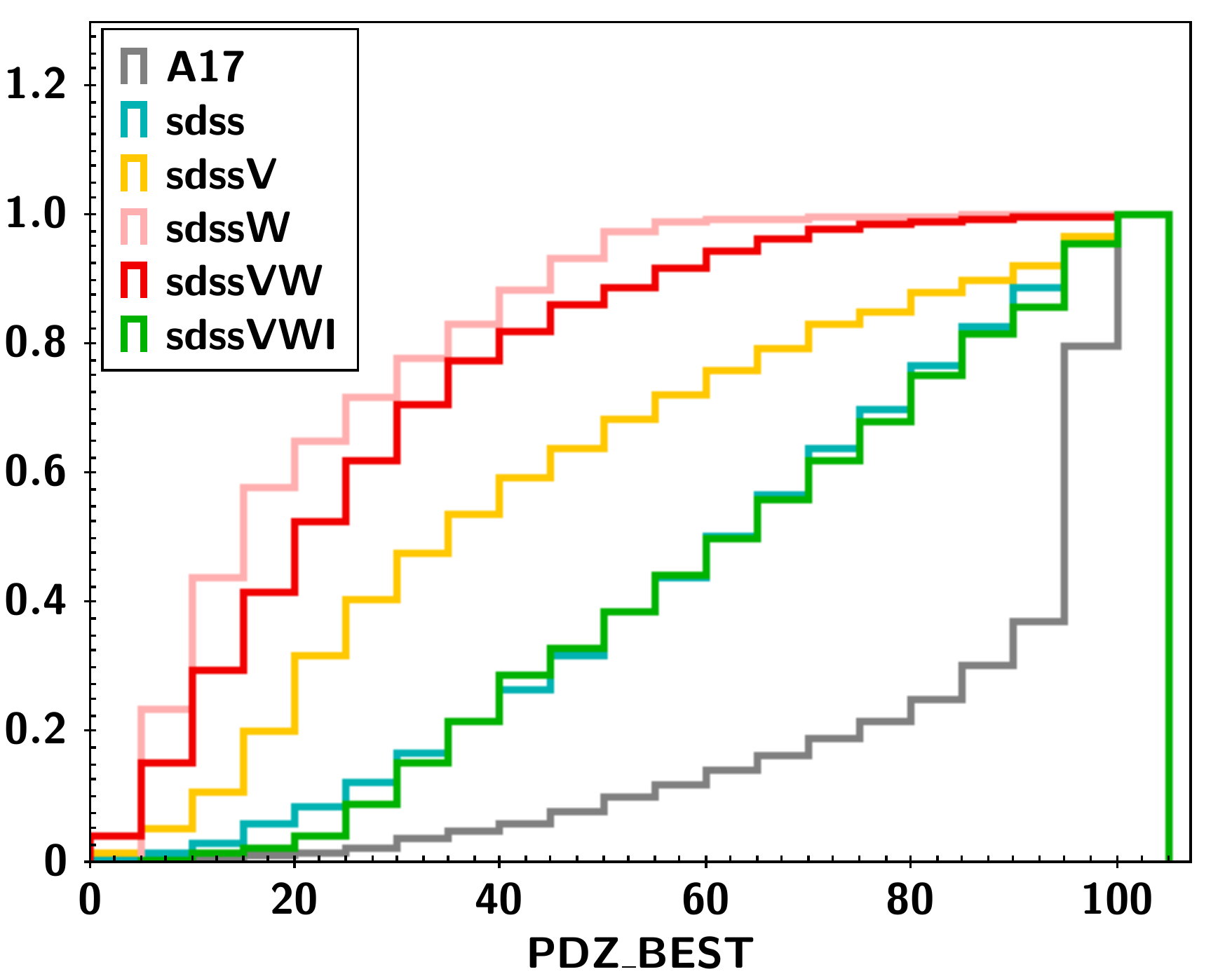}
\includegraphics[width=.49\textwidth]{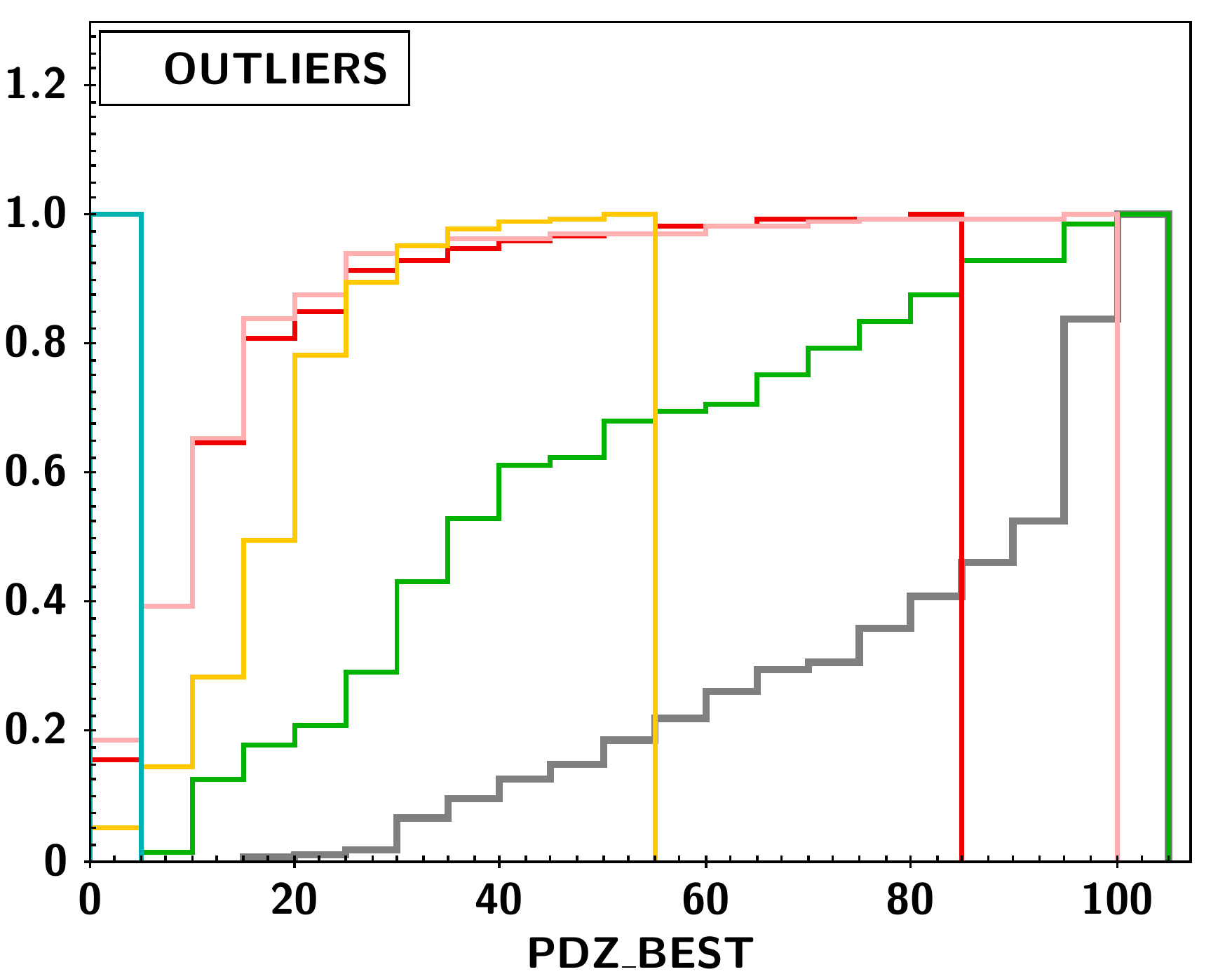}
\caption{PDZ\_BEST cumulative distribution for the entire spectroscopic sample (left) and for the outliers in the respective sub-samples used in this work (right), compared with the results from \citetalias{Ananna2017}. The comparison is missing sdssVI and sdssI, only for brevity. While the majority of the sources have a high PDZ\_BEST (e.g., larger than $85$\%) in \citetalias{Ananna2017}, METAPHOR is much more conservative and only a handful of sources reach such value. Surprisingly, LePhare assigns a high reliability to the photo-z (high PDZ\_BEST) also to the outliers, with more that $50$\% of the outliers having PDZ\_BEST> $80$.}\label{fig:histopdz}
\end{figure*}
%%%%%%%%%%%%%%%%%%%%%%%%%%%%%%%%%%%%%%%%%%%%%%
It is well known that SED-fitting algorithms tend to underestimate the error associated to the photo-z \citep{Dahlen13}. But what about the 
PDZ?\\
The PDZ is a standard product of SED-fitting algorithms and can be used in assessing how reliable a photo-z and consequently fitted SEDs are. 
Larger the PDZ, more secure the redshifts. In addition, the PDZ is now routinely used in Luminosity Functions 
\citep[e.g.,][to quote just a few]{Buchner15, Miyaji15,Fotopoulou16}.
The computation of the PDZ is a novelty in Machine Learning and here we want to test the reliability of the PDZ computed using METAPHOR 
with respect to the PDZ computed in \citetalias{Ananna2017} with LePhare.

For the analysis described below, it is important to understand how the PDZ is represented in the two methods. For LePhare, the range covered is predefined by the user. 
In \citetalias{Ananna2017} the photo-z was searched between redshift  $0$ and $7$, in bins of $0.01$ up to redshift $6$. Between 
redshift $6$ and $7$ the bin sizes are set to $0.02$. Thus,  for each source we have a file of $651$ bins. For each source, the  higher PDZ is 
normalised to $1$. It can happen that the PDZ has non-zero values only in a limited range of bins, when the photo-z solution is well defined.
In METAPHOR, although the range covered could be predefined by the user, it is recommended to set it within the extremes of the zspec distribution 
of the training sample. This  because METAPHOR, having an empirical method as internal engine, does not produce results outside the 
limits of zspec. Therefore, in this case it has been set to $[0, 5.5]$ and since only one size of bins is allowed, we choose $0.01$. It means that for each source we have a file of $551$ bins. 
As for LePhare, also with METAPHOR, if the solution is well defined, the PDZ has non-zero values only in a limited range of bins.

As discussed in \cite{Amaro18}, a unique and universal method to evaluate the PDZ reliability is extremely difficult to find. One value often used is: 
$PDZ\_BEST= \int_{z_{best}-0.1(1+z_{best})}^{z_{best}+0.1(1+z_{best})} PDZ(z)dz$ (cf. Le Phare, \citealt{Ilbert06}, documentation\footnote{\url{http://www.cfht.hawaii.edu/~arnouts/LEPHARE/DOWNLOAD/lephare_doc.pdf}}).
Here, in order to compare the results, we calculated $PDZ\_BEST$ also for METAPHOR.
For illustrative purpose, Fig.~\ref{fig:overlap} shows an example of PDZ computed with METAPHOR and compares it with the PDZ computed with LePHare for the same random sources in the catalogue.

First, we tested how often the spectroscopic redshift is close to the peak of the PDZ. This is somewhat similar to what is done in \cite{Dahlen13}, where the fraction of sources with spectroscopic redshift is within $1$ or $3$ $\sigma$ error from the photo-z. 
The analysis is  shown in Table~\ref{tabPDFclass}, where we split the cases in classes, with the following meaning: \protect\newline
1 - \zspec\ is in the bin including the PDZ peak; \protect\newline
2 - \zspec\ is in the bin beside to one including the PDZ peak; \protect\newline
3 - \zspec\ is in a bin for which the value of PDZ is zero, but it is in a bin beside to the one containing the PDZ peak; \protect\newline
4 - \zspec\ is in a bin for which the value of PDZ is more than zero (i.e., it contains classes $1$ and $2$); \protect\newline
5 - \zspec\ is in a bin for which the value of PDZ is zero, but beside to a bin for which the value of PDZ is not zero; \protect\newline
6 - \zspec\ is in a bin for which the value of PDZ is zero (contains classes $3$ and $5$).
\\

For this test, we have considered only the sources with zspec and for which the photometry in SDSS, VHS, WISE and IRAC is always available. In this way the two methods got access to the same information.
From Table~\ref{tabPDFclass}, it seems that METAPHOR is superior to the PDZ produced by LePhare in all the classes considered for the 
comparison. However, it is worth to keep in mind that the numbers presented in table are depending on the bin sizes chosen for the two 
experiments. We did not experiment with different bin sizes.

Second,  in Fig.~\ref{fig:histopdz} we looked at the PDZ\_BEST cumulative distribution for the sub-samples and compared with the results from \citetalias{Ananna2017}, 
separating the sources with reliable photo-z (left panel) from the outliers (right panel).

Overall, METAPHOR is more conservative;
while with  LePhare, less than $10$\% of the sources with spectroscopic redshift have PDZ\_BEST$<$50\%, the distribution for the sub-samples with PDZ computed by METAPHOR is completely different ($32$\%, $64$\%, $98$\%, $86$\%, $33$\% for \textit{sdss}, \textit{sdssV}, \textit{sdssW}, \textit{sdssVW}, \textit{sdssVWI}, respectively).  

The overconfidence of the PDZ computed by LePhare is even more evident in the right panel of Fig.~\ref{fig:histopdz}, where we focus on 
the outliers. The various PDZ computed with METAPHOR are low for the large majority of the samples, with no PDZ computed at all for the outliers 
in sdssVWI. In contrast, LePhare assigns a PDZ\_BEST $>80$\% to the $42$\% of its outliers.

This indicates that the PDZ computed via SED-fitting, using only a limited number of broad band photometric points, cannot be compared 
with the accuracy obtained in deep fields with many bands, where the precision of the PDZ is well tested \citep[e.g., XMM-COSMOS]{Lusso10}.
Here, with less than a dozen of photometric points, the PDZ\_BEST is very high for the majority of the outliers. One can argue that the 
outliers are not more than 20\% overall. However, neither the outliers nor their photo-z are randomly distributed. 
Therefore, a too \textit{optimistic PDZ} is more affecting certain regions of the mag/redshift/luminosity parameter space. 
In contrast, ML does not oversell the photo-z for the outliers and the PDZ\_BEST remain very low in general.

%%%%%%%%%%%%%%%%%%%%%%%%%%%%%%%%%%%%%%%%%%%%%%%%%%%%%%%%%%
\begin{table}
\centering
\begin{tabular}{crr}
Class& METAPHOR & \citetalias{Ananna2017}\\
\hline
1 & 5.4\%&2.8\%\\
2 & 14.0\%&7.8\%\\
3 & 0.0\%&0.1\%\\
4 & 99.9\%&96.8\%\\
5 & 0.0\%&0.3\%\\
6 & 0.1\%&3.2\%\\\hline
\end{tabular}
\caption{Distribution of sources with spectroscopy from \citetalias{Ananna2017} and in the sdssVWI sample, among classes defined as a function of the position of the redshifts within the PDZ.} \label{tabPDFclass}
\end{table}
%%%%%%%%%%%%%%%%%%%%%%%%%%%%%%%%%%%%%%%%%%%%%%%%%%%%%%%%%%

\section{Summary \& Conclusions}\label{sec:Conclusions} 
With the launch of eROSITA, we are facing the challenge of computing the photo-z  in a reliable and fast manner for about $3$ million sources distributed in the entire sky, with multi-wavelength data that are non homogeneous in depth and wavelength coverage. Given that photo-z computed with ML are becoming the trend in cosmological surveys involving normal galaxies, we wanted to test whether this is a viable solution also for AGN. With this purpose in mind we have used the multi-wavelength catalogue of the counterparts to the X-ray selected sources detected in Stripe 82X \citep[]{LaMassa16}, presented in \citetalias{Ananna2017}. As machine learning method we have tested MLPQNA \citep{Brescia13}.
The catalogue containing the photo-z computed for this paper, is released here. An excerpt of it is shown in Appendix~\ref{app:catalog}.

We have compared our photo-z with those computed via SED-fitting with LePhare \citep[]{Arnouts99,Ilbert06}, presented in 
\citetalias{Ananna2017}.
The main conclusions drawn from the comparison are:

\begin{itemize}
\item Before computing the photo-z with MLPQNA and machine learning in general, a feature analysis should be performed. 
Besides the obvious advantage of reducing the parameter space under analysis (i.e., to minimize the regression problem complexity), the feature selection mechanism aims also at finding an exhaustive sub-set of features, able to maintain a high photo-z prediction accuracy, but avoiding any information redundancy occurrence and thus degeneracy in the results.
The best features are not fixed but change, depending on the dataset available.
When only magnitudes are considered, the K band is by far the most important feature. The reason is easily understood keeping in mind the SED of a galaxy. The rest-frame K band indicates the knee of the SED and this clear feature can be used to determine the redshift.
When also colours are available, the importance of single band photometry is drastically reduced and the first four colours in Fig.~\ref{fig:fsmagcolhisto} represents more than $63$\% of the key feature relevance.
\item In this particular experiment, with a rich training sample able to represent the parent population and with optical, NIR and MIR 
data available, the accuracy of the photo-z computed with MLPQNA is comparable to the accuracy obtained via SED-fitting with LePhare. Comparable 
are also the fractions of outliers. When limiting the sample to the bright X-ray sources that eROSITA will detect (by comparing 
Table~\ref{tab:compl} and Fig.~\ref{fig:photo-z0}), MLPQNA performs slightly better (smaller fraction of outliers and absence of systematics). This is reassuring, given that these types of data are or will be available for the entire sky, thanks to the increasing depth of WISE \citep{Schlafly19} and the planned launch of SpherEx \citep{Dore18}.

\item Once that the training sample is large enough, the remaining limitation for ML algorithms  is in the treatment of missing data. 
Currently, in our experiment, about 35\% are lacking a photo-z for this reason. However, the use of fluxes instead of magnitudes and adding the 
photometric errors to the list of parameters \citep[using e.g.,][]{Reis19} should solve the issue.

\item As for photo-z computed via SED-fitting, the  accuracy of photo-z can improve for AGN with small photometric error and for which information on the morphology (e.g., point-like or extended) is available.

\item As mentioned already in literature, the completeness of the training sample is extremely crucial for ML algorithms. We can confirm that this is even more the case for AGN.
It can be seen by comparing the accuracy obtained with the spectroscopic sample, available when this work started, with the accuracy obtained on the $257$ new, fainter  sources from \cite{LaMassa19}. The accuracy decreases and the fraction of outliers increases noticeably. The problem is not limited to machine learning, as also the results from \citetalias{Ananna2017} worsened for this new sample. As discussed in \citetalias{Ananna2017}, the selection of the templates is based on the available spectroscopic sample. It means that irrespective of the method that will be used for computing the photo-z for eROSITA, we will need to make sure to define a training sample gathering from all the various surveys, (CDFS, COSMOS etc.), all the sources with X-ray properties typical of those that eROSITA will detect. This is similar to what \cite{Ruiz18} has done for 3XMM.

\item The PDZ computed by METAPHOR is reliable and tends to be in general more conservative than the one computed by LePhare.  Even in the best case of complete multi-wavelength coverage only for very few sources, the PDZ from METAPHOR is high. This is contrary to what happens with LePhare, where a high PDZ-BEST is obtained also for sources for which the photo-z is an outlier.

While we recommended to always use the PDZ when working with photo-z, we also want to point out that in surveys covering wide areas with 
shallow data, the accuracy of the PDZ computed via SED-fitting should not be taken for granted. It is important to underline that this is not the 
case for the PDZ provided, for example, in COSMOS, where not only the photo-z are reliable, but also the type $1$/$2$/gal classification is 
mostly correct \citep[e.g.,][]{Lusso10}. The limited reliability obtained here with PDZ computed via SED-fitting is due to the photometry 
available only from broad band filters, with a complete lack of photometry from intermediate and narrow band photometry \citep[see][for a 
complete discussion]{Salvato18}. 
To be aware of the issue is important when, e.g., computing luminosity functions for sources with limited photometry. This was also noticed and pointed out in \citet{Buchner15}. There, the authors suggested a method to correct the PDZs, making them  more realistic and showing how the correction needed was a factor of two larger for the survey with photo-z computed using only broad band photometry \citep[i.e., Aegis-X;][]{Nandra15}, rather than for COSMOS \citep{Salvato11} and CDFS \citep{Hsu14}, where also intermediate and narrow band photometry was available.

\item Assuming that a large and representative spectroscopic sample can be constructed for eROSITA, the only real limitation remains the multi-wavelength coverage on the entire sky. 
Although many all-sky surveys exist (e.g., DES, SkyMapper, Pan-StARRS, VHS, AllWISE/unWISE, deep enough surveys, able to allow reliable results with ML, are still not available in all bands. 
Adopting Stripe 82X as reference, we expected that for eROSITA, at least for $2/3$ of the final sample, reliable photo-z computed with ML can be obtained. SED-fitting will continue to provide more reliable results for the rest of the cases. 
The public catalogue with all our photo-z estimations is publicly released (see Appendix~\ref{sec:appendix} for details.)

\end{itemize}

\section*{Acknowledgements}
The authors thank the anonymous referee that with specific comments improved the manuscript.
MB acknowledges the \textit{INAF PRIN-SKA 2017 program 1.05.01.88.04} and the funding from \textit{MIUR Premiale 2016: MITIC}. 
MB and GL acknowledge the H2020-MSCA-ITN-2016 SUNDIAL (\textit{SUrvey Network for Deep Imaging Analysis and Learning}), financed within the Call 
H2020-EU.1.3.1.
SC acknowledges support from the project ``Quasars at high redshift: physics and Cosmology'' financed by the ASI/INAF agreement 2017-14-H.0. 
Topcat \citep{Taylor05} and STILTS \citep{Taylor06} have been used for this work.
This material is based upon work supported by the National Science Foundation under Grant Number AST-1715512.
DAMEWARE has been used for ML experiments \citep{DAMEWARE}. C$^3$ has been used for catalogue cross-matching \citep{Riccio17}.
The SDSS Web Site is \url{http://www.sdss.org/}.
The scientific results reported in this article are based  in part on data obtained from the Chandra and XMM Data Archive.

\bibliographystyle{mnras}
\bibliography{biblio} 

\begin{thebibliography}{}
\makeatletter
\relax
\def\mn@urlcharsother{\let\do\@makeother \do\$\do\&\do\#\do\^\do\_\do\%\do\~}
\def\mn@doi{\begingroup\mn@urlcharsother \@ifnextchar [ {\mn@doi@}
  {\mn@doi@[]}}
\def\mn@doi@[#1]#2{\def\@tempa{#1}\ifx\@tempa\@empty \href
  {http://dx.doi.org/#2} {doi:#2}\else \href {http://dx.doi.org/#2} {#1}\fi
  \endgroup}
\def\mn@eprint#1#2{\mn@eprint@#1:#2::\@nil}
\def\mn@eprint@arXiv#1{\href {http://arxiv.org/abs/#1} {{\tt arXiv:#1}}}
\def\mn@eprint@dblp#1{\href {http://dblp.uni-trier.de/rec/bibtex/#1.xml}
  {dblp:#1}}
\def\mn@eprint@#1:#2:#3:#4\@nil{\def\@tempa {#1}\def\@tempb {#2}\def\@tempc
  {#3}\ifx \@tempc \@empty \let \@tempc \@tempb \let \@tempb \@tempa \fi \ifx
  \@tempb \@empty \def\@tempb {arXiv}\fi \@ifundefined
  {mn@eprint@\@tempb}{\@tempb:\@tempc}{\expandafter \expandafter \csname
  mn@eprint@\@tempb\endcsname \expandafter{\@tempc}}}

\bibitem[\protect\citeauthoryear{Aggarwal, Saini  \& Kumar}{Aggarwal
  et~al.}{2009}]{AGGARWAL200913}
Aggarwal S.~K.,  Saini L.~M.,   Kumar A.,  2009, \mn@doi [International Journal
  of Electrical Power & Energy Systems]
  {https://doi.org/10.1016/j.ijepes.2008.09.003}, 31, 13

\bibitem[\protect\citeauthoryear{Amaro et~al.,}{Amaro et~al.}{2018}]{Amaro18}
Amaro V.,  et~al., 2018, \mn@doi [Monthly Notices of the Royal Astronomical
  Society] {10.1093/mnras/sty2922}, 482, 3116

\bibitem[\protect\citeauthoryear{{Ananna} et~al.,}{{Ananna}
  et~al.}{2017}]{Ananna2017}
{Ananna} T.~T.,  et~al., 2017, \mn@doi [\apj] {10.3847/1538-4357/aa937d}, \href
  {http://adsabs.harvard.edu/abs/2017ApJ...850...66A} {850, 66}

\bibitem[\protect\citeauthoryear{{Arnouts}, {Cristiani}, {Moscardini},
  {Matarrese}, {Lucchin}, {Fontana}  \& {Giallongo}}{{Arnouts}
  et~al.}{1999}]{Arnouts99}
{Arnouts} S.,  {Cristiani} S.,  {Moscardini} L.,  {Matarrese} S.,  {Lucchin}
  F.,  {Fontana} A.,   {Giallongo} E.,  1999, \mn@doi [\mnras]
  {10.1046/j.1365-8711.1999.02978.x}, \href
  {http://adsabs.harvard.edu/abs/1999MNRAS.310..540A} {310, 540}

\bibitem[\protect\citeauthoryear{{Biviano} et~al.,}{{Biviano}
  et~al.}{2013}]{Biviano13}
{Biviano} A.,  et~al., 2013, \mn@doi [\aap] {10.1051/0004-6361/201321955},
  \href {http://adsabs.harvard.edu/abs/2013A%26A...558A...1B} {558, A1}

\bibitem[\protect\citeauthoryear{{Boller}, {Freyberg}, {Tr{\"u}mper}, {Haberl},
  {Voges}  \& {Nandra}}{{Boller} et~al.}{2016}]{Boller16}
{Boller} T.,  {Freyberg} M.~J.,  {Tr{\"u}mper} J.,  {Haberl} F.,  {Voges} W.,
  {Nandra} K.,  2016, \mn@doi [\aap] {10.1051/0004-6361/201525648}, \href
  {http://adsabs.harvard.edu/abs/2016A%26A...588A.103B} {588, A103}

\bibitem[\protect\citeauthoryear{{Bolzonella}, {Miralles}  \&
  {Pell{\'o}}}{{Bolzonella} et~al.}{2000}]{Bolzonella00}
{Bolzonella} M.,  {Miralles} J.-M.,   {Pell{\'o}} R.,  2000, \aap, \href
  {http://cdsads.u-strasbg.fr/abs/2000A%26A...363..476B} {363, 476}

\bibitem[\protect\citeauthoryear{{Bovy} et~al.,}{{Bovy} et~al.}{2012}]{Bovy12}
{Bovy} J.,  et~al., 2012, \mn@doi [\apj] {10.1088/0004-637X/749/1/41}, \href
  {http://adsabs.harvard.edu/abs/2012ApJ...749...41B} {749, 41}

\bibitem[\protect\citeauthoryear{Breiman}{Breiman}{2001}]{Breiman01}
Breiman L.,  2001, \mn@doi [Machine Learning] {10.1023/A:1010933404324}, 45, 5

\bibitem[\protect\citeauthoryear{Brescia, Cavuoti, D'Abrusco, Longo  \&
  Mercurio}{Brescia et~al.}{2013}]{Brescia13}
Brescia M.,  Cavuoti S.,  D'Abrusco R.,  Longo G.,   Mercurio A.,  2013,
  \mn@doi [Astrophysical Journal] {10.1088/0004-637X/772/2/140}, 772

\bibitem[\protect\citeauthoryear{Brescia et~al.,}{Brescia
  et~al.}{2014a}]{DAMEWARE}
Brescia M.,  et~al., 2014a, \mn@doi [Publications of the Astronomical Society
  of the Pacific] {10.1086/677725}, 126, 783

\bibitem[\protect\citeauthoryear{Brescia, Cavuoti, Longo  \&
  De~Stefano}{Brescia et~al.}{2014b}]{Brescia14a}
Brescia M.,  Cavuoti S.,  Longo G.,   De~Stefano V.,  2014b, \mn@doi [Astronomy
  and Astrophysics] {10.1051/0004-6361/201424383}, 568

\bibitem[\protect\citeauthoryear{Brescia, Cavuoti  \& Longo}{Brescia
  et~al.}{2015}]{Brescia20153893}
Brescia M.,  Cavuoti S.,   Longo G.,  2015, \mn@doi [Monthly Notices of the
  Royal Astronomical Society] {10.1093/mnras/stv854}, 450, 3893

\bibitem[\protect\citeauthoryear{Brescia, Cavuoti, Amaro, Riccio, Angora,
  Vellucci  \& Longo}{Brescia et~al.}{2018}]{Brescia18}
Brescia M.,  Cavuoti S.,  Amaro V.,  Riccio G.,  Angora G.,  Vellucci C.,
  Longo G.,  2018, \mn@doi [Communications in Computer and Information Science]
  {10.1007/978-3-319-96553-6_5}, 822, 61

\bibitem[\protect\citeauthoryear{{Buchner} et~al.,}{{Buchner}
  et~al.}{2015}]{Buchner15}
{Buchner} J.,  et~al., 2015, \mn@doi [\apj] {10.1088/0004-637X/802/2/89}, \href
  {http://adsabs.harvard.edu/abs/2015ApJ...802...89B} {802, 89}

\bibitem[\protect\citeauthoryear{{Budav{\'a}ri} et~al.,}{{Budav{\'a}ri}
  et~al.}{2001}]{Budavari01}
{Budav{\'a}ri} T.,  et~al., 2001, \mn@doi [\aj] {10.1086/322131}, \href
  {http://adsabs.harvard.edu/abs/2001AJ....122.1163B} {122, 1163}

\bibitem[\protect\citeauthoryear{{Cardamone} et~al.,}{{Cardamone}
  et~al.}{2010}]{Cardamone10}
{Cardamone} C.~N.,  et~al., 2010, \mn@doi [\apjs]
  {10.1088/0067-0049/189/2/270}, \href
  {http://adsabs.harvard.edu/abs/2010ApJS..189..270C} {189, 270}

\bibitem[\protect\citeauthoryear{{Carrasco Kind} \& {Brunner}}{{Carrasco Kind}
  \& {Brunner}}{2013}]{Carrasco13}
{Carrasco Kind} M.,  {Brunner} R.~J.,  2013, in {Friedel} D.~N.,  ed.,
  Astronomical Society of the Pacific Conference Series Vol. 475, Astronomical
  Data Analysis Software and Systems XXII. p.~69

\bibitem[\protect\citeauthoryear{Cavuoti, Brescia, Longo  \& Mercurio}{Cavuoti
  et~al.}{2012}]{Cavuoti12}
Cavuoti S.,  Brescia M.,  Longo G.,   Mercurio A.,  2012, \mn@doi [Astronomy
  and Astrophysics] {10.1051/0004-6361/201219755}, 546

\bibitem[\protect\citeauthoryear{{Cavuoti}, {Brescia}, {De Stefano}  \&
  {Longo}}{{Cavuoti} et~al.}{2015a}]{Cavuoti15}
{Cavuoti} S.,  {Brescia} M.,  {De Stefano} V.,   {Longo} G.,  2015a, \mn@doi
  [Experimental Astronomy] {10.1007/s10686-015-9443-4}, \href
  {http://adsabs.harvard.edu/abs/2015ExA....39...45C} {39, 45}

\bibitem[\protect\citeauthoryear{{Cavuoti} et~al.,}{{Cavuoti}
  et~al.}{2015b}]{Cavuoti15a}
{Cavuoti} S.,  et~al., 2015b, \mn@doi [\mnras] {10.1093/mnras/stv1496}, \href
  {http://cdsads.u-strasbg.fr/abs/2015MNRAS.452.3100C} {452, 3100}

\bibitem[\protect\citeauthoryear{{Cavuoti}, {Amaro}, {Brescia}, {Vellucci},
  {Tortora}  \& {Longo}}{{Cavuoti} et~al.}{2017}]{Cavuoti17}
{Cavuoti} S.,  {Amaro} V.,  {Brescia} M.,  {Vellucci} C.,  {Tortora} C.,
  {Longo} G.,  2017, \mn@doi [\mnras] {10.1093/mnras/stw2930}, \href
  {http://adsabs.harvard.edu/abs/2017MNRAS.465.1959C} {465, 1959}

\bibitem[\protect\citeauthoryear{D'Isanto, Cavuoti, Gieseke  \&
  Polsterer}{D'Isanto et~al.}{2018}]{D'Isanto18}
D'Isanto A.,  Cavuoti S.,  Gieseke F.,   Polsterer K.,  2018, \mn@doi
  [Astronomy and Astrophysics] {10.1051/0004-6361/201833103}, 616

\bibitem[\protect\citeauthoryear{{Dahlen} et~al.,}{{Dahlen}
  et~al.}{2013}]{Dahlen13}
{Dahlen} T.,  et~al., 2013, \mn@doi [\apj] {10.1088/0004-637X/775/2/93}, \href
  {http://adsabs.harvard.edu/abs/2013ApJ...775...93D} {775, 93}

\bibitem[\protect\citeauthoryear{{Dawson} et~al.,}{{Dawson}
  et~al.}{2013}]{Dawson13}
{Dawson} K.~S.,  et~al., 2013, \mn@doi [\aj] {10.1088/0004-6256/145/1/10},
  \href {http://adsabs.harvard.edu/abs/2013AJ....145...10D} {145, 10}

\bibitem[\protect\citeauthoryear{{Delli Veneri}, {Cavuoti}, {Brescia}, {Longo}
  \& {Riccio}}{{Delli Veneri} et~al.}{2019}]{delliveneri19}
{Delli Veneri} M.,  {Cavuoti} S.,  {Brescia} M.,  {Longo} G.,   {Riccio} G.,
  2019, \mn@doi [MNRAS] {10.1093/mnras/stz856}, 486, 1377

\bibitem[\protect\citeauthoryear{{Delubac} et~al.,}{{Delubac}
  et~al.}{2017}]{Delubac17}
{Delubac} T.,  et~al., 2017, \mn@doi [\mnras] {10.1093/mnras/stw2741}, \href
  {http://adsabs.harvard.edu/abs/2017MNRAS.465.1831D} {465, 1831}

\bibitem[\protect\citeauthoryear{{Dore}}{{Dore}}{2018}]{Dore18}
{Dore} O.,  2018, in 42nd COSPAR Scientific Assembly. pp E1.16--15--18

\bibitem[\protect\citeauthoryear{{Duncan}, {Jarvis}, {Brown}  \&
  {Rottgering}}{{Duncan} et~al.}{2017}]{Duncan17}
{Duncan} K.~J.,  {Jarvis} M.~J.,  {Brown} M.~J.~I.,   {Rottgering} H.~J.~A.,
  2017, preprint (\mn@eprint {arXiv} {1712.04476})

\bibitem[\protect\citeauthoryear{{Duncan}, {Jarvis}, {Brown}  \&
  {R{\"o}ttgering}}{{Duncan} et~al.}{2018}]{Duncan18}
{Duncan} K.~J.,  {Jarvis} M.~J.,  {Brown} M.~J.~I.,   {R{\"o}ttgering}
  H.~J.~A.,  2018, \mn@doi [\mnras] {10.1093/mnras/sty940}, \href
  {http://adsabs.harvard.edu/abs/2018MNRAS.477.5177D} {477, 5177}

\bibitem[\protect\citeauthoryear{Ebrahimzadeh \& Khazaee}{Ebrahimzadeh \&
  Khazaee}{2010}]{EBRAHIMZADEH2010103}
Ebrahimzadeh A.,  Khazaee A.,  2010, \mn@doi [Measurement]
  {https://doi.org/10.1016/j.measurement.2009.07.002}, 43, 103

\bibitem[\protect\citeauthoryear{{Fliri} \& {Trujillo}}{{Fliri} \&
  {Trujillo}}{2016}]{Fliri16}
{Fliri} J.,  {Trujillo} I.,  2016, \mn@doi [\mnras] {10.1093/mnras/stv2686},
  \href {http://adsabs.harvard.edu/abs/2016MNRAS.456.1359F} {456, 1359}

\bibitem[\protect\citeauthoryear{{Fotopoulou} \& {Paltani}}{{Fotopoulou} \&
  {Paltani}}{2018}]{Fotopoulou18}
{Fotopoulou} S.,  {Paltani} S.,  2018, \mn@doi [\aap]
  {10.1051/0004-6361/201730763}, \href
  {http://adsabs.harvard.edu/abs/2018A%26A...619A..14F} {619, A14}

\bibitem[\protect\citeauthoryear{{Fotopoulou} et~al.,}{{Fotopoulou}
  et~al.}{2012}]{Fotopoulou12}
{Fotopoulou} S.,  et~al., 2012, \mn@doi [\apjs] {10.1088/0067-0049/198/1/1},
  \href {http://adsabs.harvard.edu/abs/2012ApJS..198....1F} {198, 1}

\bibitem[\protect\citeauthoryear{{Fotopoulou} et~al.,}{{Fotopoulou}
  et~al.}{2016}]{Fotopoulou16}
{Fotopoulou} S.,  et~al., 2016, \mn@doi [\aap] {10.1051/0004-6361/201424763},
  \href {https://ui.adsabs.harvard.edu/abs/2016A&A...587A.142F} {587, A142}

\bibitem[\protect\citeauthoryear{{Georgakakis} et~al.,}{{Georgakakis}
  et~al.}{2014}]{Georgakakis14}
{Georgakakis} A.,  et~al., 2014, \mn@doi [\mnras] {10.1093/mnras/stu1326},
  \href {http://adsabs.harvard.edu/abs/2014MNRAS.443.3327G} {443, 3327}

\bibitem[\protect\citeauthoryear{Gheyas \& Smith}{Gheyas \&
  Smith}{2010}]{Gheyas10}
Gheyas I.~A.,  Smith L.~S.,  2010, \mn@doi [Pattern Recognition]
  {https://doi.org/10.1016/j.patcog.2009.06.009}, 43, 5

\bibitem[\protect\citeauthoryear{Guyon \& Elisseeff}{Guyon \&
  Elisseeff}{2003}]{Guyon03}
Guyon I.,  Elisseeff A.,  2003, J. Mach. Learn. Res., 3, 1157

\bibitem[\protect\citeauthoryear{Hastie, Tibshirani  \& Friedman}{Hastie
  et~al.}{2009}]{hastie09}
Hastie T.,  Tibshirani R.,   Friedman J.,  2009, The Elements of Statistical
  Learning: Data Mining, Inference, and Prediction, Second Edition.
Springer Series in Statistics, Springer New York

\bibitem[\protect\citeauthoryear{{Hsu} et~al.,}{{Hsu} et~al.}{2014}]{Hsu14}
{Hsu} L.-T.,  et~al., 2014, \mn@doi [\apj] {10.1088/0004-637X/796/1/60}, \href
  {http://adsabs.harvard.edu/abs/2014ApJ...796...60H} {796, 60}

\bibitem[\protect\citeauthoryear{{Ilbert} et~al.,}{{Ilbert}
  et~al.}{2006}]{Ilbert06}
{Ilbert} O.,  et~al., 2006, \mn@doi [\aap] {10.1051/0004-6361:20065138}, \href
  {http://cdsads.u-strasbg.fr/abs/2006A%26A...457..841I} {457, 841}

\bibitem[\protect\citeauthoryear{{Ilbert} et~al.,}{{Ilbert}
  et~al.}{2009}]{Ilbert09}
{Ilbert} O.,  et~al., 2009, \mn@doi [\apj] {10.1088/0004-637X/690/2/1236},
  \href {http://adsabs.harvard.edu/abs/2009ApJ...690.1236I} {690, 1236}

\bibitem[\protect\citeauthoryear{{Irwin} et~al.,}{{Irwin}
  et~al.}{2004}]{Irwin04}
{Irwin} M.~J.,  et~al., 2004, in {Quinn} P.~J.,  {Bridger} A.,  eds,  \procspie
  Vol. 5493, Optimizing Scientific Return for Astronomy through Information
  Technologies. pp 411--422, \mn@doi{10.1117/12.551449}

\bibitem[\protect\citeauthoryear{{Ivezi{\'c}} et~al.,}{{Ivezi{\'c}}
  et~al.}{2019}]{Ivezic19}
{Ivezi{\'c}} {\v Z}.,  et~al., 2019, \mn@doi [\apj] {10.3847/1538-4357/ab042c},
  \href {http://adsabs.harvard.edu/abs/2019ApJ...873..111I} {873, 111}

\bibitem[\protect\citeauthoryear{Jolliffe}{Jolliffe}{2002}]{Jolliffe02}
Jolliffe I.~T.,  2002, Principal Component Analysis.
Springer

\bibitem[\protect\citeauthoryear{Kohavi}{Kohavi}{1995}]{Kohavi95astudy}
Kohavi R.,  1995, in Proceedings of the 14th International Joint Conference on
  Artificial Intelligence - Volume 2. IJCAI'95.
Morgan Kaufmann Publishers Inc., San Francisco, CA, USA, pp 1137--1143

\bibitem[\protect\citeauthoryear{Kohavi \& John}{Kohavi \&
  John}{1997}]{Kohavi97}
Kohavi R.,  John G.~H.,  1997, \mn@doi [Artificial Intelligence]
  {https://doi.org/10.1016/S0004-3702(97)00043-X}, 97, 273

\bibitem[\protect\citeauthoryear{Kursa \& Rudnicki}{Kursa \&
  Rudnicki}{2010}]{Kursa10}
Kursa M.,  Rudnicki W.,  2010, \mn@doi [Journal of Statistical Software,
  Articles] {10.18637/jss.v036.i11}, 36, 1

\bibitem[\protect\citeauthoryear{{LaMassa} et~al.,}{{LaMassa}
  et~al.}{2013a}]{LaMassa13b}
{LaMassa} S.~M.,  et~al., 2013a, \mn@doi [\mnras] {10.1093/mnras/stt553}, \href
  {http://adsabs.harvard.edu/abs/2013MNRAS.432.1351L} {432, 1351}

\bibitem[\protect\citeauthoryear{{LaMassa} et~al.,}{{LaMassa}
  et~al.}{2013b}]{LaMassa13}
{LaMassa} S.~M.,  et~al., 2013b, \mn@doi [\mnras] {10.1093/mnras/stt1837},
  \href {http://adsabs.harvard.edu/abs/2013MNRAS.436.3581L} {436, 3581}

\bibitem[\protect\citeauthoryear{{LaMassa} et~al.,}{{LaMassa}
  et~al.}{2016}]{LaMassa16}
{LaMassa} S.~M.,  et~al., 2016, \mn@doi [\apj] {10.3847/0004-637X/817/2/172},
  \href {http://adsabs.harvard.edu/abs/2016ApJ...817..172L} {817, 172}

\bibitem[\protect\citeauthoryear{{LaMassa}, {Georgakakis}, {Vivek}, {Salvato},
  {Tasnim Ananna}, {Urry}, {MacLeod}  \& {Ross}}{{LaMassa}
  et~al.}{2019}]{LaMassa19}
{LaMassa} S.~M.,  {Georgakakis} A.,  {Vivek} M.,  {Salvato} M.,  {Tasnim
  Ananna} T.,  {Urry} C.~M.,  {MacLeod} C.,   {Ross} N.,  2019, arXiv e-prints

\bibitem[\protect\citeauthoryear{{Laigle} et~al.,}{{Laigle}
  et~al.}{2016}]{Laigle16}
{Laigle} C.,  et~al., 2016, \mn@doi [\apjs] {10.3847/0067-0049/224/2/24}, \href
  {http://adsabs.harvard.edu/abs/2016ApJS..224...24L} {224, 24}

\bibitem[\protect\citeauthoryear{Lal, Chapelle, Weston  \& Elisseeff}{Lal
  et~al.}{2006}]{Lal06}
Lal T.~N.,  Chapelle O.,  Weston J.,   Elisseeff A.,  2006, Embedded Methods.
Springer Berlin Heidelberg, Berlin, Heidelberg, pp 137--165,
  \mn@doi{10.1007/978-3-540-35488-8_6}

\bibitem[\protect\citeauthoryear{{Laureijs}}{{Laureijs}}{2010}]{Laureijs2010}
{Laureijs} R.,  2010, in JENAM 2010, Joint European and National Astronomy
  Meeting. p.~166

\bibitem[\protect\citeauthoryear{{Lawrence} et~al.,}{{Lawrence}
  et~al.}{2007}]{Lawrence07}
{Lawrence} A.,  et~al., 2007, \mn@doi [\mnras]
  {10.1111/j.1365-2966.2007.12040.x}, \href
  {http://adsabs.harvard.edu/abs/2007MNRAS.379.1599L} {379, 1599}

\bibitem[\protect\citeauthoryear{{Luo} et~al.,}{{Luo} et~al.}{2010}]{Luo10}
{Luo} B.,  et~al., 2010, \mn@doi [\apjs] {10.1088/0067-0049/187/2/560}, \href
  {http://adsabs.harvard.edu/abs/2010ApJS..187..560L} {187, 560}

\bibitem[\protect\citeauthoryear{{Lusso} et~al.,}{{Lusso}
  et~al.}{2010}]{Lusso10}
{Lusso} E.,  et~al., 2010, \mn@doi [\aap] {10.1051/0004-6361/200913298}, \href
  {http://adsabs.harvard.edu/abs/2010A%26A...512A..34L} {512, A34}

\bibitem[\protect\citeauthoryear{{Marchesi} et~al.,}{{Marchesi}
  et~al.}{2016}]{Marchesi16}
{Marchesi} S.,  et~al., 2016, \mn@doi [\apj] {10.3847/0004-637X/817/1/34},
  \href {http://adsabs.harvard.edu/abs/2016ApJ...817...34M} {817, 34}

\bibitem[\protect\citeauthoryear{{Martin} et~al.,}{{Martin}
  et~al.}{2005}]{Martin05}
{Martin} D.~C.,  et~al., 2005, \mn@doi [\apjl] {10.1086/426387}, \href
  {http://adsabs.harvard.edu/abs/2005ApJ...619L...1M} {619, L1}

\bibitem[\protect\citeauthoryear{{Matute} et~al.,}{{Matute}
  et~al.}{2012}]{Matute12}
{Matute} I.,  et~al., 2012, \mn@doi [\aap] {10.1051/0004-6361/201118111}, \href
  {http://adsabs.harvard.edu/abs/2012A%26A...542A..20M} {542, A20}

\bibitem[\protect\citeauthoryear{{Merloni} et~al.,}{{Merloni}
  et~al.}{2012}]{Merloni12}
{Merloni} A.,  et~al., 2012, preprint (\mn@eprint {arXiv} {1209.3114})

\bibitem[\protect\citeauthoryear{{Meshcheryakov}, {Glazkova}, {Gerasimov}  \&
  {Mashechkin}}{{Meshcheryakov} et~al.}{2018}]{Meshcheryakov18}
{Meshcheryakov} A.~V.,  {Glazkova} V.~V.,  {Gerasimov} S.~V.,   {Mashechkin}
  I.~V.,  2018, \mn@doi [Astronomy Letters] {10.1134/S1063773718120058}, \href
  {http://adsabs.harvard.edu/abs/2018AstL...44..735M} {44, 735}

\bibitem[\protect\citeauthoryear{{Miyaji} et~al.,}{{Miyaji}
  et~al.}{2015}]{Miyaji15}
{Miyaji} T.,  et~al., 2015, \mn@doi [\apj] {10.1088/0004-637X/804/2/104}, \href
  {http://adsabs.harvard.edu/abs/2015ApJ...804..104M} {804, 104}

\bibitem[\protect\citeauthoryear{Mostafa}{Mostafa}{2010}]{MOSTAFA20106302}
Mostafa M.~M.,  2010, \mn@doi [Expert Systems with Applications]
  {https://doi.org/10.1016/j.eswa.2010.02.091}, 37, 6302

\bibitem[\protect\citeauthoryear{Mountrichas, Corral, Masoura, Georgantopoulos,
  Ruiz, Georgakakis, Carrera  \& Fotopoulou}{Mountrichas
  et~al.}{2017}]{Mountrichas17}
Mountrichas G.,  Corral A.,  Masoura V.,  Georgantopoulos I.,  Ruiz A.,
  Georgakakis A.,  Carrera F.,   Fotopoulou S.,  2017, \mn@doi [Astronomy and
  Astrophysics] {10.1051/0004-6361/201731762}, 608

\bibitem[\protect\citeauthoryear{{Nandra} et~al.,}{{Nandra}
  et~al.}{2015}]{Nandra15}
{Nandra} K.,  et~al., 2015, \mn@doi [\apjs] {10.1088/0067-0049/220/1/10}, \href
  {http://adsabs.harvard.edu/abs/2015ApJS..220...10N} {220, 10}

\bibitem[\protect\citeauthoryear{{Nicastro} et~al.,}{{Nicastro}
  et~al.}{2018}]{Nicastro18}
{Nicastro} F.,  et~al., 2018, \mn@doi [\nat] {10.1038/s41586-018-0204-1}, \href
  {http://adsabs.harvard.edu/abs/2018Natur.558..406N} {558, 406}

\bibitem[\protect\citeauthoryear{Papovich et~al.,}{Papovich
  et~al.}{2016}]{Papovich}
Papovich C.,  et~al., 2016, The Astrophysical Journal Supplement Series, 224,
  28

\bibitem[\protect\citeauthoryear{{P{\^a}ris} et~al.,}{{P{\^a}ris}
  et~al.}{2018}]{Paris18}
{P{\^a}ris} I.,  et~al., 2018, \mn@doi [\aap] {10.1051/0004-6361/201732445},
  \href {http://adsabs.harvard.edu/abs/2018A%26A...613A..51P} {613, A51}

\bibitem[\protect\citeauthoryear{Pearson}{Pearson}{2010}]{Pearson2010}
Pearson K.,  2010, \mn@doi [The London, Edinburgh, and Dublin Philosophical
  Magazine and Journal of Science] {10.1080/14786440109462720}, 2, 559

\bibitem[\protect\citeauthoryear{{Polsterer}, {Gieseke}, {Igel}  \&
  {Goto}}{{Polsterer} et~al.}{2014}]{Polsterer14}
{Polsterer} K.~L.,  {Gieseke} F.,  {Igel} C.,   {Goto} T.,  2014, in {Manset}
  N.,  {Forshay} P.,  eds,  Astronomical Society of the Pacific Conference
  Series Vol. 485, Astronomical Data Analysis Software and Systems XXIII.
  p.~425

\bibitem[\protect\citeauthoryear{{Reis}, {Baron}  \& {Shahaf}}{{Reis}
  et~al.}{2019}]{Reis19}
{Reis} I.,  {Baron} D.,   {Shahaf} S.,  2019, \mn@doi [\aj]
  {10.3847/1538-3881/aaf101}, \href
  {http://adsabs.harvard.edu/abs/2019AJ....157...16R} {157, 16}

\bibitem[\protect\citeauthoryear{{Riccio}, {Brescia}, {Cavuoti}, {Mercurio},
  {di Giorgio}  \& {Molinari}}{{Riccio} et~al.}{2017}]{Riccio17}
{Riccio} G.,  {Brescia} M.,  {Cavuoti} S.,  {Mercurio} A.,  {di Giorgio} A.~M.,
    {Molinari} S.,  2017, \mn@doi [\pasp] {10.1088/1538-3873/129/972/024005},
  \href {http://adsabs.harvard.edu/abs/2017PASP..129b4005R} {129, 024005}

\bibitem[\protect\citeauthoryear{Rosenblatt}{Rosenblatt}{1962}]{Rosenblatt61}
Rosenblatt F.,  1962, Principles of neurodynamics; perceptrons and the theory
  of brain mechanisms..
Spartan Books, Washington

\bibitem[\protect\citeauthoryear{{Ruiz}, {Corral}, {Mountrichas}  \&
  {Georgantopoulos}}{{Ruiz} et~al.}{2018}]{Ruiz18}
{Ruiz} A.,  {Corral} A.,  {Mountrichas} G.,   {Georgantopoulos} I.,  2018,
  \mn@doi [\aap] {10.1051/0004-6361/201833117}, \href
  {http://adsabs.harvard.edu/abs/2018A%26A...618A..52R} {618, A52}

\bibitem[\protect\citeauthoryear{{Sadeh}, {Abdalla}  \& {Lahav}}{{Sadeh}
  et~al.}{2016}]{Sadeh16}
{Sadeh} I.,  {Abdalla} F.~B.,   {Lahav} O.,  2016, \mn@doi [\pasp]
  {10.1088/1538-3873/128/968/104502}, \href
  {http://adsabs.harvard.edu/abs/2016PASP..128j4502S} {128, 104502}

\bibitem[\protect\citeauthoryear{{Salvato} et~al.,}{{Salvato}
  et~al.}{2009}]{Salvato09}
{Salvato} M.,  et~al., 2009, \mn@doi [\apj] {10.1088/0004-637X/690/2/1250},
  \href {http://adsabs.harvard.edu/abs/2009ApJ...690.1250S} {690, 1250}

\bibitem[\protect\citeauthoryear{{Salvato} et~al.,}{{Salvato}
  et~al.}{2011}]{Salvato11}
{Salvato} M.,  et~al., 2011, \mn@doi [\apj] {10.1088/0004-637X/742/2/61}, \href
  {http://adsabs.harvard.edu/abs/2011ApJ...742...61S} {742, 61}

\bibitem[\protect\citeauthoryear{{Salvato}, {Ilbert}  \& {Hoyle}}{{Salvato}
  et~al.}{2018}]{Salvato18}
{Salvato} M.,  {Ilbert} O.,   {Hoyle} B.,  2018, \mn@doi [Nature Astronomy]
  {10.1038/s41550-018-0478-0}

\bibitem[\protect\citeauthoryear{{Schlafly}, {Meisner}  \& {Green}}{{Schlafly}
  et~al.}{2019}]{Schlafly19}
{Schlafly} E.~F.,  {Meisner} A.~M.,   {Green} G.~M.,  2019, \mn@doi [\apjs]
  {10.3847/1538-4365/aafbea}, \href
  {http://adsabs.harvard.edu/abs/2019ApJS..240...30S} {240, 30}

\bibitem[\protect\citeauthoryear{{Tanaka}}{{Tanaka}}{2015}]{Tanaka15}
{Tanaka} M.,  2015, \mn@doi [\apj] {10.1088/0004-637X/801/1/20}, \href
  {http://cdsads.u-strasbg.fr/abs/2015ApJ...801...20T} {801, 20}

\bibitem[\protect\citeauthoryear{{Tangaro} et~al.,}{{Tangaro}
  et~al.}{2015}]{Tangaro:2015}
{Tangaro} S.,  et~al., 2015, \mn@doi [Computational and Mathematical Methods in
  Medicine, vol. 2015, 814104] {10.1155/2015/814104}

\bibitem[\protect\citeauthoryear{{Taylor}}{{Taylor}}{2005}]{Taylor05}
{Taylor} M.~B.,  2005, in {Shopbell} P.,  {Britton} M.,   {Ebert} R.,  eds,
  Astronomical Society of the Pacific Conference Series Vol. 347, Astronomical
  Data Analysis Software and Systems XIV. p.~29

\bibitem[\protect\citeauthoryear{{Taylor}}{{Taylor}}{2006}]{Taylor06}
{Taylor} M.~B.,  2006, in {Gabriel} C.,  {Arviset} C.,  {Ponz} D.,   {Enrique}
  S.,  eds,  Astronomical Society of the Pacific Conference Series Vol. 351,
  Astronomical Data Analysis Software and Systems XV. p.~666

\bibitem[\protect\citeauthoryear{Tibshirani}{Tibshirani}{2013}]{tibshirani12}
Tibshirani R.~J.,  2013, \mn@doi [Electron. J. Statist.] {10.1214/13-EJS815},
  7, 1456

\bibitem[\protect\citeauthoryear{Timlin et~al.,}{Timlin et~al.}{2016}]{Timlin}
Timlin J.~D.,  et~al., 2016, The Astrophysical Journal Supplement Series, 225,
  1

\bibitem[\protect\citeauthoryear{{Vanzella} et~al.,}{{Vanzella}
  et~al.}{2004}]{Vanzella04}
{Vanzella} E.,  et~al., 2004, \mn@doi [\aap] {10.1051/0004-6361:20040176},
  \href {http://adsabs.harvard.edu/abs/2004A%26A...423..761V} {423, 761}

\bibitem[\protect\citeauthoryear{{Voges} et~al.,}{{Voges}
  et~al.}{1999}]{Voges99}
{Voges} W.,  et~al., 1999, \aap, \href
  {http://adsabs.harvard.edu/abs/1999A%26A...349..389V} {349, 389}

\bibitem[\protect\citeauthoryear{{Wright} et~al.,}{{Wright}
  et~al.}{2010}]{Wright10}
{Wright} E.~L.,  et~al., 2010, \mn@doi [\aj] {10.1088/0004-6256/140/6/1868},
  \href {http://adsabs.harvard.edu/abs/2010AJ....140.1868W} {140, 1868}

\bibitem[\protect\citeauthoryear{Zare, Pourghasemi, Vafakhah  \& Pradhan}{Zare
  et~al.}{2013}]{Zare13}
Zare M.,  Pourghasemi H.~R.,  Vafakhah M.,   Pradhan B.,  2013, \mn@doi
  [Arabian Journal of Geosciences] {10.1007/s12517-012-0610-x}, 6, 2873

\bibitem[\protect\citeauthoryear{{de Jong} et~al.,}{{de Jong}
  et~al.}{2017}]{de_Jong:17}
{de Jong} J.~T.~A.,  et~al., 2017, \mn@doi [\aap]
  {10.1051/0004-6361/201730747}, \href
  {http://adsabs.harvard.edu/abs/2017arXiv170302991D} {604, A134}

\makeatother
\end{thebibliography}
%\clearpage

\section*{SUPPORTING INFORMATION}
Supplementary data are available at \href{https://academic.oup.com/mnras/article/489/1/663/5549519#supplementary-data}{MNRAS} online.\\

\textbf{PHOTOZ ML for PAPER 2019JUL24.dat}\\

\noindent Please note: Oxford University Press is not responsible for the
content or functionality of any supporting materials supplied by
the authors. Any queries (other than missing material) should be
directed to the corresponding author for the article
\appendix
\section{Photo-z comparison and public catalogue}\label{sec:appendix} 

%%%%%%%%%%%%%%%%%%%%%%%%%%%%%%%%%%%%%%%%%%%%%%%
\begin{table*}
\centering
\begin{tabular}{lccccccccccc} 
REC\_{NO} &  CTP\_{RA} & CTP\_{DEC} & sdssVWI & sdssVW & sdssVI & sdssW &  sdssI & sdssV & sdss & MLPQNA\_{merged}  \\
 \hline
  \hline
    1  & 0.980191   &  0.2046045  &  1.06487 &  1.11284  &  -99.0 &  1.0867  &  -99.0 &  1.16123 &  1.10542 &  1.06487\\
    2  & 0.9812111  &  0.1268089  &  1.19282 &  0.955105 &  -99.0 &  0.98219 &  -99.0 &  1.10901 &  1.14375 &  1.19282\\
    3  & 0.9939954  &  0.0559024  &  -99.0   &  -99.0    &  -99.0 &  -99.0   &  -99.0 &  -99.0   &  1.44947 &  1.44947\\
    4  & 1.0114936  &  0.1948067  &  -99.0   &  0.194302 &  -99.0 &  0.10898 &  -99.0 &  0.25149 &  0.16807 &  0.194302\\
    5  & 1.0116661  &  0.1634697  &  -99.0   &  -99.0    &  -99.0 &  -99.0   &  -99.0 &  -99.0   &  -99.0   &  -99.0\\
\end{tabular}
\caption{Example of contents of the photo-z catalogue made publicly available.}
\label{tab:catalogue}
\end{table*}

\begin{figure*}
\centering
\includegraphics[width=\textwidth,trim={1.5cm 1.5cm 2cm 2cm},clip]{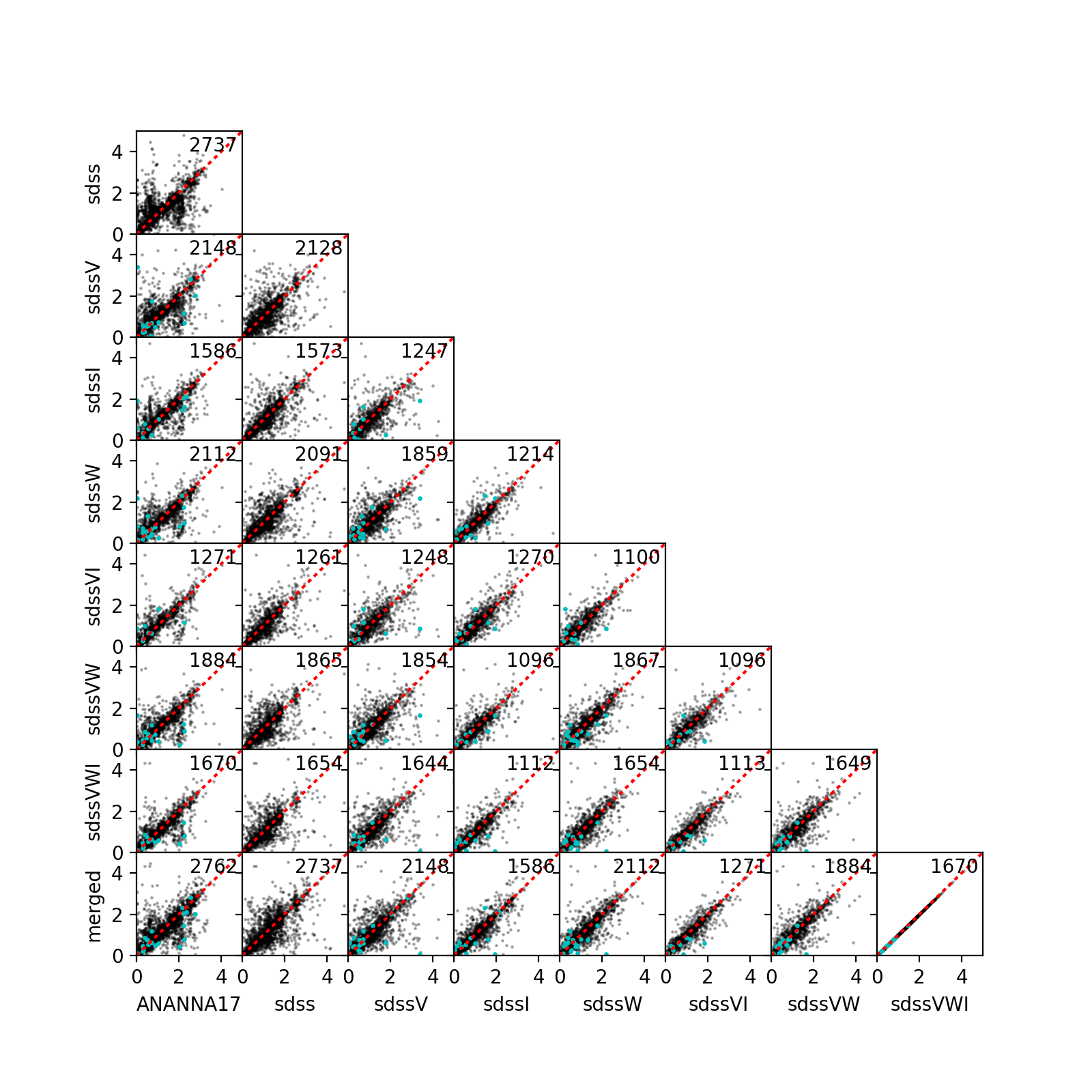}

\caption{Comparison between photo-z computed via SED fitting (\citetalias{Ananna2017}) and MLPQNA for the sample for which spectroscopic 
information is available. The cyan points indicate the sources for which the redshift could be computed only after considering supplementary 
photometry in addition to SDSS.}
\label{fig:photo-z3allWzspec}
\end{figure*}
%%%%%%%%%%%%%%%%%%%%%%%%%%%%%%%%%%%%%%%%%%
\begin{figure*}
\centering
\includegraphics[width=\textwidth,trim={1.5cm 1.5cm 2cm 2cm},clip]{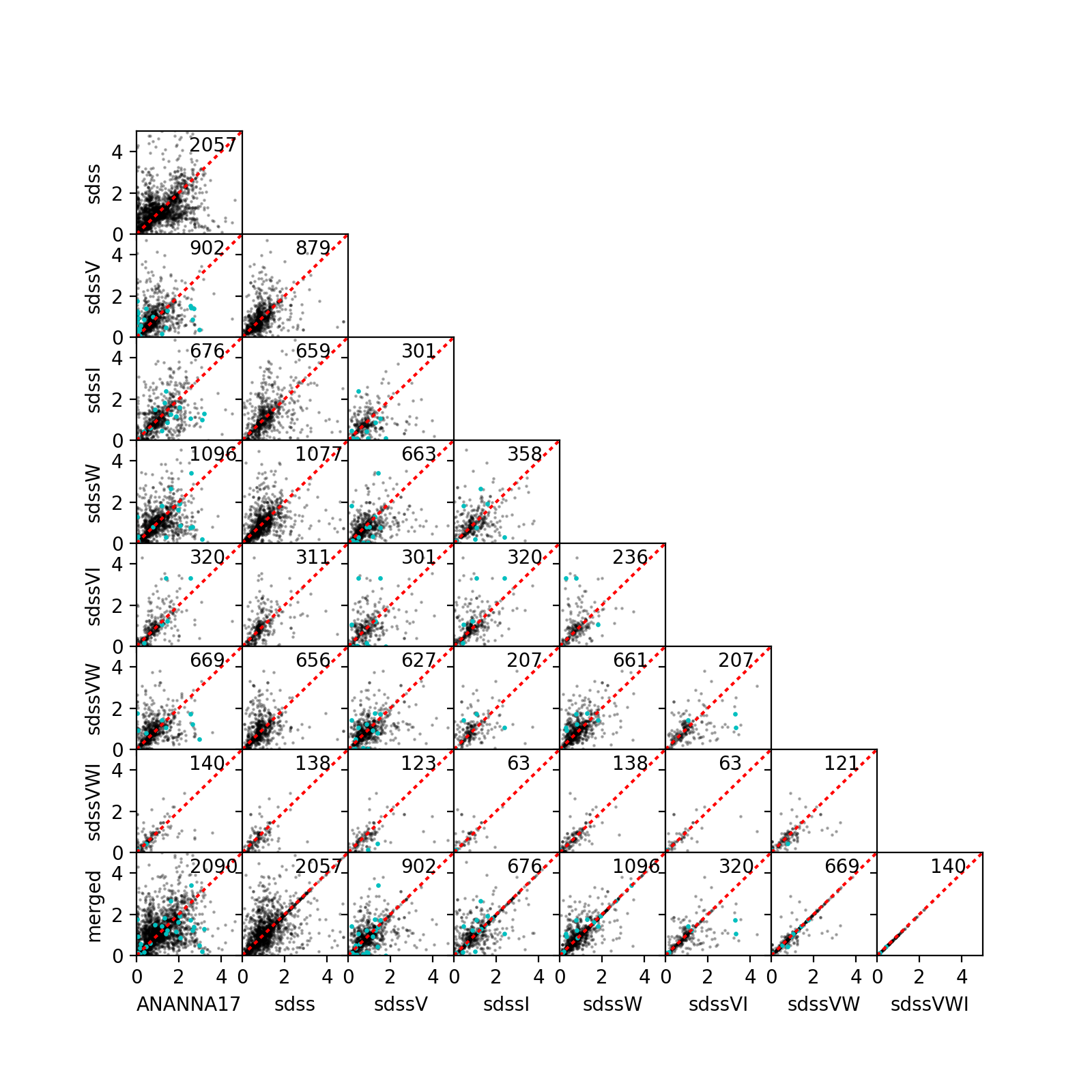}

\caption{Comparison between photo-z computed via SED fitting (\citetalias{Ananna2017}) and MLPQNA for the sample for which spectroscopic 
information is not available. The cyan points indicate the sources for which the redshift could be computed only after considering 
supplementary photometry in addition to SDSS.}
\label{fig:photo-z3allNozspec}
\end{figure*}

%%%%%%%%%%%%%%%%%%%%%%%%%%%%%%%%%%%%%%%%%%
\subsection{Catalogue Release} \label{app:catalog}
The produced catalogue of photo-z, obtained by different cross-matches among available surveys as well as their final best combination, 
is made publicly available at \href{https://academic.oup.com/mnras/article/489/1/663/5549519#supplementary-data}{MNRAS} online. A sample of the internal structure is shown in Table~\ref{tab:catalogue}. 
The catalogue is indexed on the first column, which can be used to retrieve all other information about spectroscopic redshifts and X-ray 
source counterparts, by cross-matching this catalogue with the one referred in \citetalias{Ananna2017}. The other columns, from left to right, 
are respectively, RA and DEC of optical counterparts, followed by the photo-z estimations obtained by all discussed combinations of surveys, 
i.e. SDSS, VHS, IRAC and WISE. The last column is related to the best photo-z obtained from all the previous combinations, as explained in 
the main text.

% Don't change these lines
\bsp	% typesetting comment
\label{lastpage}
\end{document}